\documentclass[preprint,12pt]{elsarticle}

\usepackage{amssymb}
\usepackage{amsmath}
\usepackage{booktabs}
\usepackage{multirow}
\usepackage{hyperref}
\usepackage{xcolor}

\hypersetup{colorlinks=true, linkcolor=blue, citecolor=blue, urlcolor=blue}

\newcommand{\rd}{r_\mathrm{d}}
\newcommand{\DM}{D_\mathrm{M}}
\renewcommand{\DH}{D_\mathrm{H}}
\newcommand{\DV}{D_\mathrm{V}}
\newcommand{\Rperp}{\mathcal{R}_\perp}

\newcommand{\LCDM}{\Lambda\mathrm{CDM}}
\newcommand{\tBAO}{\theta_\mathrm{BAO}}
\newcommand{\zeff}{z_\mathrm{eff}}

\journal{Physics of the Dark Universe}

\begin{document}

\begin{frontmatter}

\title{On the origin of the BAOtr--DESI tension}

\author[ioannina]{Ioannis Pantos\corref{cor1}}
\author[ioannina]{Leandros Perivolaropoulos}
\ead{i.pantos@uoi.gr; leandros@uoi.gr}

\cortext[cor1]{Corresponding author}
\address[ioannina]{Department of Physics, University of Ioannina,
GR-45110 Ioannina, Greece}

\begin{abstract}
The fiducial-independent angular/transverse BAO dataset, obtained from two-point angular correlation functions in thin redshift shells (hereafter BAOtr), systematically
prefers smaller comoving distance ratios $\DM/\rd$ than the DESI DR2
three-dimensional BAO measurements at $z \lesssim 0.65$, driving dataset-dependent CPL dark-energy inferences and, consequently, conflicting conclusions about the Hubble tension. We investigate whether this disagreement
can be attributed to the $\LCDM$ fiducial assumed in the 3D BAO
pipeline, or resolved within the CPL dark-energy parametrisation.
We show that the published 3D BAO distances are fiducial-independent
by construction, with residual effects at $\lesssim 0.3\%$ ---
negligible against the 10--18\% BAOtr uncertainties.  We then scan
the CPL parameter space with $\Omega_m$ and $H_0$ jointly determined
at each $(w_0, w_a)$ by the Planck $\theta_*$ constraint and
optimisation against the DESI data.  Two complementary tests are
performed: a direct comparison of each DESI-optimized model with the
BAOtr data, and an $\alpha$-interpolation test that anchors the
prediction to the DESI measurements and asks whether the CPL
backbone provides a better interpolation between them than $\LCDM$.
Both reveal an inescapable trade-off: models that fit DESI well
($\chi^2_{\rm DESI} \lesssim 5$) yield
$\chi^2_{\rm BAOtr} \gtrsim 42$, while reducing the BAOtr tension
to $\chi^2_{\rm BAOtr} \sim 37$ requires $\chi^2_{\rm DESI} \gtrsim 8$. No CMB-consistent CPL model fits
both datasets simultaneously.  The direct, model-independent
comparison at $z = 0.510$ --- where BAOtr and DESI disagree by
$3.7\sigma$ (data-versus-data)--- sets an irreducible tension floor that no smooth
modification of $\DM(z)$ can remove.  These conclusions are robust
across analysis methods, extrapolation schemes, and substitution of
SDSS for DESI as the reference dataset.  The remaining explanations
are observational systematics --- most plausibly in the BAOtr
measurements --- or new physics beyond CPL.
\end{abstract}

\end{frontmatter}

%% ═══════════════════════════════════════════════════════════════════
\section{Introduction}
\label{sec:intro}
%% ═══════════════════════════════════════════════════════════════════

%% ~3 pages
%% - BAO as standard ruler, DESI DR2, CPL preference (1 para)
%% - Xu et al. finding: BAOtr-DESI tension, dataset dependence (1 para)
%% - Fiducial question: methodological asymmetry (1 para)
%% - Three hypotheses (brief list)
%% - This paper's contribution (1 para)
%% - Outline (1 para)

Baryon acoustic oscillations (BAO) imprint a characteristic comoving
scale --- the sound horizon at the baryon drag epoch,
$\rd \approx 147$\,Mpc --- on the matter distribution
\cite{Eisenstein:1998,Eisenstein:2005,Meiksin:1999,Blake:2003,Seo:2003}.
Observed as a peak in the galaxy two-point correlation function, this
scale serves as a standard ruler that constrains the comoving angular
diameter distance $\DM(z)$ and the Hubble parameter $H(z)$ across
cosmic time.  BAO measurements have matured from the first detections
in SDSS \cite{Eisenstein:2005} and 2dFGRS \cite{Cole:2005}, through
the BOSS and eBOSS programmes \cite{Alam:2017,Alam:2021}, to the
multi-tracer campaign of the Dark Energy Spectroscopic Instrument
(DESI) \cite{DESIinstrument:2022,Adame:2025,AbdulKarim:2025}.  The
DESI second data release (DR2) \cite{AbdulKarim:2025}, spanning
$0.1 < z < 4.2$ with seven tracer samples, reports a $2.8\sigma$
preference for evolving dark energy in the Chevallier--Polarski--Linder
(CPL) parametrisation \cite{Chevallier:2001,Linder:2003},
$w(a) = w_0 + w_a(1-a)$, when combined with Planck CMB data
and Pantheon+ supernovae
\cite{Planck:2020} --- strengthening the earlier $2.5\sigma$ hint from
DR1 \cite{Adame:2025}. This result has stimulated extensive follow-up
analyses exploring its robustness and physical implications
\cite{Giare:2024,Cortes:2024,Wolf:2024,Giarè:2025}, adding to the
broad literature on cosmological tensions
\cite{Perivolaropoulos:2021jda,Pantos:2026koc,Tsagas:2025pxi}.

In a related direction, Afroz and
Mukherjee~\cite{Afroz:2025zot} tested the internal consistency of
the DESI DR2 BAO and Pantheon+ datasets via the distance-duality
relation and reported a redshift-dependent signal that, when
accounted for in the likelihood, suppresses the preference for
dynamical dark energy --- suggesting that part of the CPL signal
from DESI+SNIa combinations may arise from dataset inconsistency.

The methodological asymmetry between 3D and BAOtr has been
studied in a model-independent framework by Gómez-Valent,
Favale, Migliaccio and Sen~\cite{GomezValent:2024}. Using
Gaussian-process reconstructions of $H(z)$ and the supernova
absolute magnitude $M(z)$ from cosmic chronometers and BAO, they
showed that the two datasets require qualitatively different
late-time phenomenologies: anisotropic BAO combined with a local
$H_0$ prior demand a phantom-like enhancement of $H(z)$ at
$z \lesssim 0.2$ and a transition in $M(z)$, while BAOtr
are compatible with SH0ES without such features. A model- and
calibrator-independent quantification of the same tension was
subsequently carried out by Favale, Gómez-Valent and
Migliaccio~\cite{Favale:2024}, who used SNIa as a redshift
interpolator and the Etherington relation to compare the two
datasets without calibrating $r_d$ or $M$. They found tensions
of $\sim 2$--$2.5\sigma$ with Pantheon+ and $\sim 4.6\sigma$
with DES~Y5, confirming that the disagreement is not an
artefact of any specific parametrisation or sound-horizon
calibration. The dataset-dependent CPL inferences were further documented by
Gómez-Valent and Solà Peracaula~\cite{GomezValent:2025}, whose
CPL fits with either 2D or 3D BAO anticipate the pattern later
reported by Xu~et~al.

A critical examination of the dataset dependence of this preference
has been carried out by Xu~et~al.\ \cite{Xu:2026}, who combine CMB
anisotropies and lensing with four suites of late-time probes:
DESI~DR2 BAO, the SDSS-IV consensus \cite{Alam:2021}, a
transverse/angular BAO compilation (BAOtr) \cite{Nunes:2020}, and the
Cepheid-calibrated PantheonPlus supernovae (PP\&SH0ES)
\cite{Scolnic:2022,Brout:2022,Riess:2022}.  Their central finding is
that the CPL inferences are strongly dataset-dependent: combinations
involving 3D BAO (DESI or SDSS) prefer low $H_0$ and a distinctive
$(w_0, w_a)$ locus, while combinations involving BAOtr and/or
PP\&SH0ES shift toward higher $H_0$ and a different CPL region.  The
origin of this bifurcation is traced to a specific observational
disagreement: the BAOtr dataset systematically prefers \emph{smaller}
$\DM/\rd$ than the DESI DR2 measurements at all overlapping redshifts
$z \lesssim 0.65$, with offsets of 10--13\% at individual points.
This low-redshift mismatch propagates through the CPL parameter space
and is identified as the root cause of the divergent late-time
reconstructions.

The BAOtr--BAO\,3D disagreement is of particular significance because
the two measurement approaches differ fundamentally in their
relationship to the assumed cosmological model.  Standard 3D BAO
analyses require a \emph{fiducial cosmology} to convert observed
angular separations and redshift differences into comoving distances,
to apply the Alcock--Paczy\'{n}ski (AP) correction \cite{AP:1979},
and to construct the template correlation function against which the
dilation parameters $\alpha_\perp$ and $\alpha_\parallel$ are
measured.  Both DESI~DR2 and the SDSS-IV consensus adopt a Planck
$\LCDM$ fiducial ($\Omega_m = 0.3153$,
$H_0 = 67.36$\,km\,s$^{-1}$\,Mpc$^{-1}$,
$\rd^{\rm fid} = 147.05$\,Mpc).  By contrast, BAOtr measurements
\cite{Nunes:2020,Sanchez:2011,Carvalho:2016,Carvalho:2020,
deCarvalho:2021,Alcaniz:2017,deCarvalho:2018} extract the angular BAO
scale $\tBAO(z) = \rd/\DM(z)$ from the two-point angular correlation
function $w(\theta)$ in thin redshift shells, without any conversion
to comoving coordinates.  The BAOtr observable is therefore
fiducial-independent by construction.  This methodological asymmetry
raises a natural question: could the BAOtr--BAO\,3D tension be a
consequence of the $\LCDM$ fiducial assumed in the 3D analyses?

Three logically distinct classes of explanation exist for the observed
tension:
\begin{enumerate}
  \item \textbf{Fiducial misspecification.}  The $\LCDM$ fiducial
    biases the published 3D BAO distances, and BAOtr reveals this
    bias.  Adopting a CPL fiducial would reduce the disagreement.
  \item \textbf{Observational systematics.}  Unidentified systematic
    effects in one or both methodologies --- projection effects,
    photometric-redshift uncertainties, or angular-mask artefacts in
    BAOtr; residual template mismatch or reconstruction biases in
    3D BAO --- produce a coherent offset.
  \item \textbf{New physics.}  The true distance--redshift relation
    differs from both $\LCDM$ and CPL in ways that the two-parameter
    CPL ansatz cannot capture.  Models with richer late-time dynamics
    --- such as $\Lambda_s$CDM
    \cite{Akarsu:2020,Akarsu:2021,Akarsu:2023}, omnipotent dark
    energy \cite{Adil:2024}, phantom-crossing scenarios
    \cite{DiValentino:2021,Escamilla:2023}, or braneworld models
    \cite{Sahni:2003} --- might provide a better description.
\end{enumerate}

In this paper we focus on hypothesis~(i) and test whether the CPL
parametrisation can reconcile the two BAO datasets.  Our analysis
proceeds in two stages.  First, we show that the published 3D BAO
distance ratios $\DM/\rd$ are fiducial-independent as a direct
consequence of the definition of the dilation parameters: the fiducial
dependence of $\alpha_\perp$ cancels identically against that of
$\DM^{\rm fid}/\rd^{\rm fid}$ in the published product.  Residual
effects (AP peak distortion, reconstruction sensitivity,
broadband-polynomial incompleteness) have been quantified by the DESI
collaboration at $\lesssim 0.1$--$0.3\%$
\cite{AbdulKarim:2025,PerezFernandez:2024} --- negligible compared to
the 10--18\% BAOtr uncertainties.  A ``rescaling'' approach that
multiplies published distances by the ratio
$\Rperp = \DM^{\rm CPL}/\DM^{\LCDM}$ therefore overcorrects by a
factor of 20--50.  Second, we scan the CPL parameter space with
$\Omega_m$ adjusted at each $(w_0, w_a)$ to satisfy the Planck
angular scale constraint $\theta_*$, and compare each CMB-consistent
model with both the DESI and BAOtr data.  We find that no CPL model
fits both datasets simultaneously: the $\chi^2_{\rm DESI}$ vs
$\chi^2_{\rm BAOtr}$ trade-off traces an anti-correlation across the
entire parameter space, with no model reaching the lower-left
(good-fit) corner.  The direct comparison at $z = 0.510$ ---
where BAOtr and DESI measure $\DM/\rd$ at the same redshift and
disagree by $3.7\sigma$ (data-versus-data) --- sets an irreducible floor that no
smooth model change can remove.  These results are confirmed by an
independent $\alpha$-interpolation analysis that anchors predictions
to the DESI data points, and are robust across extrapolation schemes,
BGS anchor inclusion, and substitution of SDSS for DESI.  The
fiducial-misspecification hypothesis is therefore ruled out, and the
CPL framework cannot reconcile the two BAO datasets.  The remaining
possibilities are observational systematics --- most plausibly in the
BAOtr measurements --- or new physics beyond CPL.  This is
consistent with recent independent findings that BAO miscalibrations
are unlikely to rescue late-time solutions to the Hubble tension
\cite{Pedrotti:2025ccw}.

The paper is organised as follows.  Section~\ref{sec:fiducial}
establishes the fiducial independence of published 3D BAO distances
and diagnoses the rescaling overcorrection.
Section~\ref{sec:data} presents the three BAO datasets.
Section~\ref{sec:method} describes the CPL models with
CMB-consistent $\Omega_m$ and the two tension statistics.
Section~\ref{sec:results} presents the baseline tension, the CPL
scan, robustness tests, and the consistency between methods.
Section~\ref{sec:discussion} discusses the remaining explanations
and the connection to the Hubble tension.
Section~\ref{sec:conclusions} summarises our conclusions.

%% ═══════════════════════════════════════════════════════════════════

\section{Fiducial independence of published BAO distances}
\label{sec:fiducial}
%% ═══════════════════════════════════════════════════════════════════

\subsection{The 3D BAO pipeline}
\label{sec:pipeline}

The standard 3D BAO analysis pipeline, as implemented in both DESI
\cite{AbdulKarim:2025} and SDSS/BOSS/eBOSS
\cite{Alam:2017,Alam:2021}, transforms raw galaxy positions into
measurements of $\DM/\rd$ and $\DH/\rd$ through four steps, each with
a distinct relationship to the assumed fiducial cosmology.  We trace
the fiducial dependence through each step and show that it cancels
identically in the published product.

\medskip
\noindent\textbf{Step~1: Coordinate conversion.}
The observables for each galaxy pair --- angular separation
$\Delta\theta$ and redshift difference $\Delta z$ --- are converted
to comoving separations using a fiducial cosmology:
\begin{equation}
  s_\perp = \DM^{\rm fid}(z)\cdot\Delta\theta\,,\qquad
  s_\parallel = \DH^{\rm fid}(z)\cdot\Delta z\,,
  \label{eq:coord_conv}
\end{equation}
where $\DH^{\rm fid} \equiv c/H^{\rm fid}(z)$.  Both DESI~DR2 and
SDSS-IV use the Planck 2018 $\LCDM$ best fit ($\Omega_m = 0.3153$,
$H_0 = 67.36$\,km\,s$^{-1}$\,Mpc$^{-1}$,
$\rd^{\rm fid} = 147.05$\,Mpc) \cite{Planck:2020}.  If the fiducial
does not match the true cosmology, this mapping introduces the
Alcock--Paczy\'{n}ski (AP) distortion \cite{AP:1979}: an anisotropic,
smooth, scale-independent rescaling of the clustering pattern.

\medskip
\noindent\textbf{Step~2: Correlation function and reconstruction.}
The pipeline computes the anisotropic two-point correlation function
$\xi(s_\perp, s_\parallel)$ using the Landy--Szalay estimator
\cite{LandySzalay:1993}, typically after a reconstruction procedure
\cite{Eisenstein:2007rec,Padmanabhan:2012} that partially reverses
non-linear displacements and sharpens the BAO peak.  Because the
coordinate conversion uses fiducial rather than true distances, the
BAO ring appears not at $s = \rd$ but at
\begin{equation}
  s_{\rm BAO}^\perp
  = \DM^{\rm fid}(z)\cdot\frac{\rd}{\DM^{\rm true}(z)}\,,\qquad
  s_{\rm BAO}^\parallel
  = \DH^{\rm fid}(z)\cdot\frac{\rd}{\DH^{\rm true}(z)}\,.
  \label{eq:bao_peak_fid}
\end{equation}

\medskip
\noindent\textbf{Step~3: Template fitting.}
Because the fiducial cosmology generally differs from the true one,
the observed peak position (Eq.~\ref{eq:bao_peak_fid}) does not
coincide with the template prediction.  This mismatch is absorbed by
two free dilation parameters, $\alpha_\perp$ and $\alpha_\parallel$,
that independently rescale the template coordinates in the transverse
and line-of-sight directions.  A parametric model is fitted to $\xi$
\cite{Xu:2012template,Anderson:2014,Beutler:2017,Ross:2017}:
\begin{equation}
  \xi_{\rm model}(s_\perp, s_\parallel)
  = B(s)\;\xi_{\rm template}
    \!\left(\alpha_\perp\, s_\perp,\,
            \alpha_\parallel\, s_\parallel\right)
  + A(s)\,,
  \label{eq:template_fit}
\end{equation}
where $\xi_{\rm template}$ is computed in the fiducial cosmology (BAO
peak at $s = \rd^{\rm fid}$), $\alpha_\perp$ and $\alpha_\parallel$
are dilation parameters, and $B(s)$ and $A(s)$ are broadband nuisance
functions --- typically low-order polynomials in $1/s$ --- that absorb
all smooth distortions of $\xi$, including the AP effect, galaxy-bias
uncertainties, non-linear mode coupling, and window-function effects.
These smooth terms cannot absorb a shift of the sharp BAO peak (width
$\sim$\!10--20\,$h^{-1}$\,Mpc), so $\alpha_\perp$ and
$\alpha_\parallel$ are determined by the peak position alone.

In the rescaled template, the peak appears where
$\alpha_\perp\, s_\perp = \rd^{\rm fid}$, i.e.\ at
$s_\perp = \rd^{\rm fid}/\alpha_\perp$.  Matching to the data peak
(Eq.~\ref{eq:bao_peak_fid}) gives
\begin{equation}
  \alpha_\perp
  = \frac{\rd^{\rm fid} \cdot \DM^{\rm true}(z)}
         {\DM^{\rm fid}(z) \cdot \rd}
  = \frac{\DM^{\rm true}(z)/\rd}
         {\DM^{\rm fid}(z)/\rd^{\rm fid}}\,,
  \label{eq:alpha_solved}
\end{equation}
the standard definition (Eq.~(1) of Ref.~\cite{AbdulKarim:2025}).
Note the explicit fiducial dependence in the denominator.  An
analogous expression holds for $\alpha_\parallel$.

\medskip
\noindent\textbf{Step~4: Published distances and the fiducial
cancellation.}
The pipeline reports
\begin{equation}
  \left(\frac{\DM}{\rd}\right)^{\!\rm pub}
  = \alpha_\perp \cdot \frac{\DM^{\rm fid}}{\rd^{\rm fid}}\,.
  \label{eq:published}
\end{equation}
Substituting Eq.~\eqref{eq:alpha_solved}:
\begin{equation}
  \left(\frac{\DM}{\rd}\right)^{\!\rm pub}
  = \frac{\DM^{\rm true}/\rd}
         {\DM^{\rm fid}/\rd^{\rm fid}}
    \cdot\frac{\DM^{\rm fid}}{\rd^{\rm fid}}
  = \frac{\DM^{\rm true}}{\rd}\,.
  \label{eq:cancellation}
\end{equation}
The fiducial dependence of $\alpha_\perp$ (the denominator in
Eq.~\ref{eq:alpha_solved}) cancels identically against the prefactor
$\DM^{\rm fid}/\rd^{\rm fid}$ in Eq.~\eqref{eq:published}, leaving a
result that depends only on the true cosmology.  This cancellation is
a direct algebraic consequence of the definition of $\alpha_\perp$ and
holds for \emph{any} pair of fiducial cosmologies sharing the same
$\rd$.  It extends straightforwardly to the radial distance
$(\DH/\rd)^{\rm pub}$ and to the isotropic combination
$(\DV/\rd)^{\rm pub}$.  For models that change $\rd$ (e.g., early
dark energy \cite{Poulin:2019,Karwal:2016}), the published ratio
still equals $\DM^{\rm true}/\rd$; the fiducial $\rd^{\rm fid}$
cancels regardless of whether it matches the true value.

As a numerical illustration, the DESI LRG1 measurement at
$\zeff = 0.510$ gives $(\DM/\rd)^{\rm pub} = 13.588 \pm 0.167$.
Switching the fiducial from Planck $\LCDM$ to the CPL model
$(w_0, w_a) = (-0.42, -1.75)$ changes $\alpha_\perp$ by $+6.3\%$ and
$\DM^{\rm fid}/\rd$ by $-5.9\%$; the two shifts cancel exactly in
the product, which remains $13.588$.

\medskip
\noindent\textbf{Residual effects.}
The cancellation of Eq.~\eqref{eq:cancellation} is exact within the
idealised fitting model of Eq.~\eqref{eq:template_fit}.  In practice,
three effects introduce a residual fiducial sensitivity:
(i)~anisotropic distortion of the BAO peak shape from the AP effect,
which the isotropic broadband polynomials cannot fully absorb;
(ii)~sub-optimal reconstruction under a misspecified fiducial, leaving
slightly larger non-linear broadening; and (iii)~broadband-polynomial
incompleteness at finite truncation order.  The DESI collaboration has
quantified all three through mock-catalogue tests with deliberately
misspecified fiducials
\cite{AbdulKarim:2025,KP4:2024,PerezFernandez:2024}, finding a net
residual bias $\delta\alpha/\alpha \lesssim 0.1$--$0.3\%$ for
fiducials within $\sim$\!5\% of the truth.  At $z = 0.5$, this
corresponds to a bias in $\DM/\rd$ of at most $\sim$\!0.04\,units,
compared to BAOtr uncertainties of $\sigma \approx 0.4$--$0.9$\,units.
A detailed characterisation of these residual effects --- including
their dependence on the degree of fiducial mismatch and on the
broadband parametrisation --- is beyond the scope of this work but
would be valuable for future high-precision BAOtr comparisons.

\subsection{Why the rescaling approach overcorrects}
\label{sec:rescaling}

A natural but incorrect approach to testing the fiducial hypothesis
would multiply the published $(\DM/\rd)^{\rm pub}$ by the ratio
$\Rperp(z) = \DM^{\rm CPL}(z)/\DM^{\LCDM}(z)$ to ``correct'' for the
$\LCDM$ fiducial.  Since $(\DM/\rd)^{\rm pub} = \DM^{\rm true}/\rd$
is already fiducial-independent (Eq.~\ref{eq:cancellation}), this
produces
\begin{equation}
  \left(\frac{\DM}{\rd}\right)_{\!\rm rescaled}
  = \frac{\DM^{\rm true}}{\rd}\times\Rperp(z)\,,
  \label{eq:rescaled}
\end{equation}
which is the true distance multiplied by a spurious model-dependent
factor.  For the CPL models preferred by DESI+CMB analyses
($w_0 > -1$, $w_a < 0$), $\Rperp < 1$ at $z \lesssim 1$, so the
rescaling systematically \emph{shrinks} the DESI predictions ---
moving them toward BAOtr and producing an artificial tension
reduction of $\Delta\chi^2 \sim 40$--$50$.
Table~\ref{tab:overcorrection} quantifies this overcorrection.

\begin{table}[ht]
\centering
\caption{Overcorrection from the rescaling approach at representative
DESI redshifts, using the CMB+DESI CPL model
$(w_0 = -0.42,\;w_a = -1.75)$ \cite{Xu:2026}.  The applied shift
(3--6\%) exceeds the mock-validated residual fiducial bias
($\lesssim 0.3\%$) by a factor of 20--50.
$\Rperp(z) = \DM^{\rm CPL}(z)/\DM^{\LCDM}(z)$ is evaluated
numerically at each redshift using the stated CPL parameters with
Planck~2018 background values ($\Omega_m = 0.3153$,
$H_0 = 67.36$\,km\,s$^{-1}$\,Mpc$^{-1}$).}
\label{tab:overcorrection}
\setlength{\tabcolsep}{5pt}
\footnotesize
\begin{tabular}{ccccccc}
\toprule
$z$ & $(\DM/\rd)^{\rm pub}$ & $\Rperp$ & Rescaled
    & Shift (\%) & True residual (\%) & Overcorrection \\
\midrule
0.295 & 7.942$^a$  & 0.942 & 7.482  & $-5.8$ & $\lesssim 0.3$ & $\gtrsim 19\times$ \\
0.510 & 13.588     & 0.941 & 12.791 & $-5.9$ & $\lesssim 0.3$ & $\gtrsim 20\times$ \\
0.706 & 17.351     & 0.941 & 16.334 & $-5.9$ & $\lesssim 0.2$ & $\gtrsim 29\times$ \\
0.934 & 21.576     & 0.945 & 20.398 & $-5.5$ & $\lesssim 0.2$ & $\gtrsim 27\times$ \\
1.321 & 27.601     & 0.954 & 26.337 & $-4.6$ & $\lesssim 0.1$ & $\gtrsim 46\times$ \\
2.330 & 38.988     & 0.969 & 37.794 & $-3.1$ & $\lesssim 0.1$ & $\gtrsim 31\times$ \\
\bottomrule
\multicolumn{7}{l}{\footnotesize $^a$BGS $\DV/\rd$; $\Rperp$ here
denotes $\mathcal{R}_V = \DV^{\rm CPL}/\DV^{\LCDM}$.  The
$\lesssim 0.3\%$ residual is validated} \\
\multicolumn{7}{l}{\footnotesize \phantom{$^a$}by DESI mock tests
for both anisotropic and isotropic fits
\cite{AbdulKarim:2025,PerezFernandez:2024}.}
\end{tabular}
\end{table}

A useful diagnostic for any analysis claiming to correct 3D BAO
distances for fiducial effects: if the applied correction
$|f(z)-1| \times 100\%$ exceeds the mock-validated residual
($\lesssim 0.3\%$) by orders of magnitude, it is almost certainly
overcorrecting.  The correct procedure is not to modify the published
$\DM/\rd$ (which are fiducial-independent measurements of the true
distance), but to change the cosmological model used to
\emph{predict} $\DM/\rd$ and compare that prediction with both
datasets --- the approach we adopt in Section~\ref{sec:method}.

\subsection{Angular BAO: fiducial-free by construction}
\label{sec:angular}

The transversal (angular) BAO method extracts the BAO scale from the
two-point angular correlation function $w(\theta)$ of galaxies in
thin, disjoint redshift shells
\cite{Sanchez:2011,Carnero:2012,Sanchez:2014}.  Within each shell,
$w(\theta)$ is computed from angular pair counts using the
Landy--Szalay estimator \cite{LandySzalay:1993}; the BAO peak
position $\tBAO$ is identified by fitting a template to $w(\theta)$
\cite{Sanchez:2011,Carvalho:2016,Carvalho:2020}.  No fiducial
cosmology enters at any stage: the pair separation is a directly
observed angle, and the conversion to comoving distance,
\begin{equation}
  \frac{\DM(z)}{\rd} = \frac{180^\circ/\pi}{\tBAO(z)}\,,
  \label{eq:theta_to_DM}
\end{equation}
is exact for a flat universe and requires no assumed parameters.

The BAOtr compilation used in this work \cite{Nunes:2020} assembles
15 measurements from four analyses of SDSS data (DR7--DR12): one
low-redshift point from blue galaxies \cite{deCarvalho:2021}
($z = 0.110$), two from photometric LRGs \cite{Alcaniz:2017}
($z = 0.235, 0.365$), eleven from spectroscopic LRGs
\cite{Carvalho:2016,Carvalho:2020} ($z = 0.450$--$0.650$), and one
from photometric quasars \cite{deCarvalho:2018} ($z = 2.225$).  The
measurements span $0.110 \leq z \leq 2.225$ with the densest coverage
at $0.45 \lesssim z \lesssim 0.65$, and are statistically independent
by construction (disjoint shells).

Several systematic effects limit the precision of BAOtr and
could contribute to the observed tension.  Projection effects from
finite shell width ($\Delta z = 0.02$--$0.04$) can shift $\tBAO$ at
the $\sim$\!0.5--1\% level \cite{Sanchez:2011,Nishimichi:2007,Crocce:2011}.
Photometric redshift errors ($\sigma_z/(1+z) \sim 0.02$--$0.04$)
broaden the effective shell and amplify projection biases
\cite{Padmanabhan:2007,Ross:2011}.  All 15 measurements share the
SDSS imaging footprint and its angular systematics (seeing, Galactic
extinction, stellar contamination)
\cite{Ross:2011,Ho:2012,Leistedt:2013}, so a coherent angular
systematic would bias all redshift bins in the same direction ---
mimicking a cosmological signal.  These effects have not been
characterised at the sub-percent level required by current 3D BAO
comparisons, and we return to them in
Section~\ref{sec:explanations}.

%% ═══════════════════════════════════════════════════════════════════
\section{Data}
\label{sec:data}
%% ═══════════════════════════════════════════════════════════════════

\subsection{DESI DR2 BAO}
\label{sec:desi_data}

The DESI DR2 BAO measurements \cite{AbdulKarim:2025} are listed in
Table~\ref{tab:DESI}.  We use the 13 independent measurements from
seven tracer samples spanning $0.295 \leq \zeff \leq 2.330$.  The
BGS sample at $\zeff = 0.295$ provides only the isotropic distance
$\DV/\rd = 7.942 \pm 0.075$; the remaining six tracers provide both
$\DM/\rd$ and $\DH/\rd$.  For the BAOtr comparison, which constrains
only $\DM/\rd$, the six anisotropic $\DM/\rd$ values (rows~2, 4, 6,
8, 10, 12) serve as anchor points, with the BGS measurement entering
directly as a $\DV/\rd$ constraint.  Fractional uncertainties on
$\DM/\rd$ range from 0.7\% (LRG3+ELG1) to 2.5\% (QSO), making the
DESI values 3--15 times more precise than BAOtr at comparable
redshifts.  As established in Section~\ref{sec:pipeline}, all values
are fiducial-independent.

\begin{table}[ht]
\centering
\caption{DESI DR2 BAO distance measurements \cite{AbdulKarim:2025}:
13 independent measurements from seven tracers.  Distances are in
units of $\rd$.  The correlation coefficient
$\rho_{M,H} \equiv \rho(\DM/\rd,\,\DH/\rd)$ is listed for
anisotropic tracers.  Fiducial:
$\rd^{\rm fid} = 147.05$\,Mpc.}
\label{tab:DESI}
\setlength{\tabcolsep}{3.2pt}
\footnotesize
\begin{tabular}{clcccccl}
\toprule
\# & Tracer & $z$-range & $\zeff$ & Observable
   & Value $\pm\,\sigma$ & $\rho_{M,H}$ & Type \\
\midrule
 1 & BGS
   & 0.1--0.4 & 0.295
   & $\DV/\rd$ & $7.942 \pm 0.075$ & ---
   & Bright galaxies \\[3pt]
 2 & LRG1
   & 0.4--0.6 & 0.510
   & $\DM/\rd$ & $13.588 \pm 0.167$ & \multirow{2}{*}{$-0.459$}
   & \multirow{2}{*}{Lum.\ red galaxies} \\
 3 & LRG1
   & 0.4--0.6 & 0.510
   & $\DH/\rd$ & $21.863 \pm 0.425$ & & \\[3pt]
 4 & LRG2
   & 0.6--0.8 & 0.706
   & $\DM/\rd$ & $17.351 \pm 0.177$ & \multirow{2}{*}{$-0.404$}
   & \multirow{2}{*}{Lum.\ red galaxies} \\
 5 & LRG2
   & 0.6--0.8 & 0.706
   & $\DH/\rd$ & $19.455 \pm 0.330$ & & \\[3pt]
 6 & LRG3+ELG1
   & 0.8--1.1 & 0.934
   & $\DM/\rd$ & $21.576 \pm 0.152$ & \multirow{2}{*}{$-0.416$}
   & \multirow{2}{*}{LRGs + ELGs} \\
 7 & LRG3+ELG1
   & 0.8--1.1 & 0.934
   & $\DH/\rd$ & $17.641 \pm 0.193$ & & \\[3pt]
 8 & ELG2
   & 1.1--1.6 & 1.321
   & $\DM/\rd$ & $27.601 \pm 0.318$ & \multirow{2}{*}{$-0.434$}
   & \multirow{2}{*}{Emission-line gal.} \\
 9 & ELG2
   & 1.1--1.6 & 1.321
   & $\DH/\rd$ & $14.176 \pm 0.221$ & & \\[3pt]
10 & QSO
   & 0.8--2.1 & 1.484
   & $\DM/\rd$ & $30.512 \pm 0.760$ & \multirow{2}{*}{$-0.500$}
   & \multirow{2}{*}{Quasars} \\
11 & QSO
   & 0.8--2.1 & 1.484
   & $\DH/\rd$ & $12.817 \pm 0.516$ & & \\[3pt]
12 & Ly$\alpha$
   & 1.8--4.2 & 2.330
   & $\DM/\rd$ & $38.988 \pm 0.531$ & \multirow{2}{*}{$-0.431$}
   & \multirow{2}{*}{Ly$\alpha$ forest} \\
13 & Ly$\alpha$
   & 1.8--4.2 & 2.330
   & $\DH/\rd$ & $8.632 \pm 0.101$ & & \\
\bottomrule
\end{tabular}
\end{table}

\subsection{SDSS-IV consensus BAO}
\label{sec:sdss_data}

As a cross-check, we use the SDSS-IV (BOSS + eBOSS) consensus
compilation \cite{Alam:2021}, listed in Table~\ref{tab:SDSS}.  Five
$\DM/\rd$ measurements (marked $\star$) span
$\zeff = 0.380$--$2.334$.  Compared to DESI, SDSS provides an
additional anchor at $z = 0.380$ within the critical tension region,
but lacks a BGS-equivalent below $z = 0.380$.  At overlapping
redshifts the two surveys agree within 1.5--4.5\%: at $z = 0.510$, SDSS
gives $13.36 \pm 0.21$ vs DESI $13.588 \pm 0.167$ ($0.85\sigma$);
at $z \approx 2.33$, SDSS gives $37.30 \pm 1.70$ vs DESI
$38.988 \pm 0.531$ ($0.95\sigma$).  This consistency confirms that the
3D BAO distance scale is robust across independent surveys.  For the
Ly$\alpha$ sector, we use the cross-correlation measurement
(row~13) following standard practice \cite{Alam:2021,Xu:2026}.

\begin{table}[ht]
\centering
\caption{SDSS-IV (BOSS + eBOSS) consensus BAO measurements
\cite{Alam:2021}.  Rows marked $\star$ are the five $\DM/\rd$
anchors used in our cross-check analysis.}
\label{tab:SDSS}
\setlength{\tabcolsep}{3.2pt}
\footnotesize
\begin{tabular}{clcccl}
\toprule
\# & Sample & $\zeff$ & Observable
   & Value $\pm\,\sigma$ & Note \\
\midrule
 1 & MGS & 0.150
   & $\DV/\rd$ & $4.47 \pm 0.17$
   & Isotropic$^{a}$ \\[3pt]
 2$^{\star}$ & BOSS Galaxy & 0.380
   & $\DM/\rd$ & $10.23 \pm 0.17$
   & $^{b}$ \\
 3 & BOSS Galaxy & 0.380
   & $\DH/\rd$ & $25.00 \pm 0.76$
   & $^{b}$ \\[3pt]
 4$^{\star}$ & BOSS Galaxy & 0.510
   & $\DM/\rd$ & $13.36 \pm 0.21$
   & $^{b}$ \\
 5 & BOSS Galaxy & 0.510
   & $\DH/\rd$ & $22.33 \pm 0.58$
   & $^{b}$ \\[3pt]
 6$^{\star}$ & eBOSS LRG & 0.700
   & $\DM/\rd$ & $17.86 \pm 0.33$
   & $^{c,d}$ \\
 7 & eBOSS LRG & 0.700
   & $\DH/\rd$ & $19.33 \pm 0.53$
   & $^{c,d}$ \\[3pt]
 8 & eBOSS ELG & 0.845
   & $\DV/\rd$ & $18.33 \pm 0.62$
   & Isotropic$^{e}$ \\[3pt]
 9 & eBOSS QSO & 1.480
   & $\DH/\rd$ & $13.26 \pm 0.55$
   & $^{f,g}$ \\
10$^{\star}$ & eBOSS QSO & 1.480
   & $\DM/\rd$ & $30.69 \pm 0.80$
   & $^{f,g}$ \\[3pt]
11 & Ly$\alpha$ auto & 2.330
   & $\DM/\rd$ & $37.60 \pm 1.90$
   & Corr.\ w/\ 13$^{h}$ \\
12 & Ly$\alpha$ auto & 2.330
   & $\DH/\rd$ & $8.93 \pm 0.28$
   & Corr.\ w/\ 13$^{h}$ \\[3pt]
13$^{\star}$ & Ly$\alpha\!\times\!$QSO & 2.334
   & $\DM/\rd$ & $37.30 \pm 1.70$
   & Cross-corr.$^{h}$ \\
14 & Ly$\alpha\!\times\!$QSO & 2.334
   & $\DH/\rd$ & $9.08 \pm 0.34$
   & $^{h}$ \\
\bottomrule
\multicolumn{6}{l}{\footnotesize
$^{a}$\,\cite{Ross:2015};\;
$^{b}$\,\cite{Alam:2017};\;
$^{c}$\,\cite{Bautista:2021};\;
$^{d}$\,\cite{GilMarin:2020};\;
$^{e}$\,\cite{deMattia:2021};\;
$^{f}$\,\cite{Neveux:2020};\;
$^{g}$\,\cite{Hou:2021};\;
$^{h}$\,\cite{duMas:2020}.}
\end{tabular}
\end{table}

\subsection{BAOtr compilation}
\label{sec:baotr_data}

Table~\ref{tab:BAOtr} lists the 15 BAOtr measurements from
Ref.~\cite{Nunes:2020}, with derived $\DM/\rd$ values computed via
Eq.~\eqref{eq:theta_to_DM}.  Fractional uncertainties range from
2.5\% ($z = 0.235$) to 17.5\% ($z = 2.225$), roughly an order of
magnitude larger than the DESI measurements at comparable redshifts.
The data show non-monotonic scatter in $\DM/\rd$ at
$z = 0.45$--$0.65$ (e.g., $\DM/\rd$ drops from 12.01 at $z = 0.450$
to 11.41 at $z = 0.470$), consistent with the 4--7\% statistical
uncertainties but cautioning against over-interpreting individual
points.  Only one BAOtr measurement exactly coincides with a DESI
anchor: at $z = 0.510$, BAOtr gives $\DM/\rd = 11.912 \pm 0.421$
while DESI gives $13.588 \pm 0.167$ --- a direct, interpolation-free
difference of $3.7\sigma$. An independent cross-check comes from DES Type Ia supernovae:
reconstructing $\DM/\rd$ from DES SNe at the same effective redshift
$z = 0.510$, assuming a Planck $\rd$, yields
$\DM/\rd = 13.32 \pm 0.15$ (E.\ Ó Colgáin, private communication,
to appear in the revised version of Ref.~\cite{LopezHernandez:2025xxx}),
fully consistent with DESI and discrepant with BAOtr at $\sim 3.2\sigma$.
A SH0ES-like $\rd$ would push this value higher, reinforcing the
agreement with DESI. The 15 measurements are drawn from four
distinct SDSS analyses using different galaxy selections and
template-fitting methods, providing some protection against a single
analysis-specific bias, but all share the SDSS imaging footprint and
its angular systematics.

\begin{table}[ht]
\centering
\caption{Transversal (angular) BAO measurements \cite{Nunes:2020}.
$\tBAO(z) = \rd/\DM(z)$ is measured from $w(\theta)$ without a
fiducial cosmology.  Derived $\DM/\rd$ values follow from
Eq.~\eqref{eq:theta_to_DM}.}
\label{tab:BAOtr}
\setlength{\tabcolsep}{4pt}
\footnotesize
\begin{tabular}{clccccl@{\hspace{6pt}}c}
\toprule
\# & $z$ & $\tBAO$ [deg] & $\sigma_\theta$ [deg]
   & $\DM/\rd$ & $\sigma_{\DM/\rd}$
   & Source & $\sigma_\theta/\tBAO$ \\
\midrule
 1 & 0.110 & 19.80 & 3.26 &  2.894 & 0.477
   & \cite{deCarvalho:2021}$^{a}$ & 16.5\% \\
 2 & 0.235 &  9.06 & 0.23 &  6.324 & 0.161
   & \cite{Alcaniz:2017}$^{b}$ & 2.5\% \\
 3 & 0.365 &  6.33 & 0.22 &  9.052 & 0.315
   & \cite{Alcaniz:2017}$^{b}$ & 3.5\% \\[2pt]
 4 & 0.450 &  4.77 & 0.17 & 12.012 & 0.428
   & \cite{Carvalho:2016}$^{c}$ & 3.6\% \\
 5 & 0.470 &  5.02 & 0.25 & 11.414 & 0.568
   & \cite{Carvalho:2016}$^{c}$ & 5.0\% \\
 6 & 0.490 &  4.99 & 0.21 & 11.483 & 0.483
   & \cite{Carvalho:2016}$^{c}$ & 4.2\% \\
 7 & 0.510 &  4.81 & 0.17 & 11.912 & 0.421
   & \cite{Carvalho:2016}$^{c}$ & 3.5\% \\
 8 & 0.530 &  4.29 & 0.30 & 13.356 & 0.934
   & \cite{Carvalho:2016}$^{c}$ & 7.0\% \\
 9 & 0.550 &  4.25 & 0.25 & 13.481 & 0.793
   & \cite{Carvalho:2016}$^{c}$ & 5.9\% \\[2pt]
10 & 0.570 & 4.62 & 0.40 & 12.402 & 1.074
   & \cite{Carvalho:2020}$^{d}$ & 8.7\% \\
11 & 0.590 & 4.37 & 0.35 & 13.111 & 1.050
   & \cite{Carvalho:2020}$^{d}$ & 8.0\% \\
12 & 0.610 & 3.86 & 0.33 & 14.843 & 1.269
   & \cite{Carvalho:2020}$^{d}$ & 8.5\% \\
13 & 0.630 & 3.88 & 0.42 & 14.767 & 1.598
   & \cite{Carvalho:2020}$^{d}$ & 10.8\% \\
14 & 0.650 & 3.54 & 0.17 & 16.185 & 0.777
   & \cite{Carvalho:2020}$^{d}$ & 4.8\% \\[2pt]
15 & 2.225 &  1.77 & 0.31 & 32.371 & 5.667
   & \cite{deCarvalho:2018}$^{e}$ & 17.5\% \\
\bottomrule
\multicolumn{8}{l}{\footnotesize
$^{a}$\,SDSS DR7 blue galaxies;\;
$^{b}$\,SDSS DR10/DR11 phot.\ LRGs;\;
$^{c}$\,SDSS DR10 LRGs;} \\
\multicolumn{8}{l}{\footnotesize
$^{d}$\,SDSS DR11 LRGs;\;
$^{e}$\,SDSS DR12 phot.\ quasars.}
\end{tabular}
\end{table}

\subsection{The BAOtr--BAO\texorpdfstring{\,}{}3D tension}
\label{sec:tension_summary}

Xu~et~al.\ \cite{Xu:2026} identify a systematic offset: BAOtr prefers
smaller $\DM/\rd$ than DESI at all overlapping redshifts
$z \lesssim 0.65$, with the discrepancy most pronounced at
$z \approx 0.35$--$0.55$ (individual offsets of $\sim$\!10--13\% in
$\DM/\rd$, corresponding to 2--4$\sigma$ per point).  The same
pattern is observed with SDSS replacing DESI.  This low-redshift
disagreement propagates into divergent CPL inferences: CMB + 3D BAO
combinations drive the fit toward $w_0 \approx -0.4$, $w_a \approx
-1.8$, with $H_0 \approx 64$\,km\,s$^{-1}$\,Mpc$^{-1}$, while
CMB + BAOtr combinations prefer $w_0 \approx -0.7$, $w_a \approx
-1.8$, with $H_0 \approx 71$--$73$\,km\,s$^{-1}$\,Mpc$^{-1}$
\cite{Xu:2026}.  The Bayesian evidence for CPL over $\LCDM$ is very
strong ($\Delta\ln\mathcal{Z} \approx 16$) when BAOtr and PP\&SH0ES
are included, but inconclusive for CMB+DESI and moderately favouring
$\LCDM$ for CMB+SDSS.  The tension is thus directly connected to the
Hubble tension: BAOtr-inclusive combinations yield $H_0$ consistent
with local measurements
\cite{Casertano:2025}, while
3D-BAO-inclusive combinations worsen the discrepancy.

As shown in Section~\ref{sec:pipeline}, the fiducial cosmology cannot
explain this disagreement: the published 3D BAO distances depend on
the fiducial at the $\lesssim 0.3\%$ level, two orders of magnitude
below the observed offsets.  The three hypotheses of
Section~\ref{sec:intro} therefore reduce to two: observational
systematics or new physics.

%% ═══════════════════════════════════════════════════════════════════

\section{Method}
\label{sec:method}
%% ═══════════════════════════════════════════════════════════════════

\subsection{CPL dark energy and CMB-consistent parameters}
\label{sec:cpl}

We adopt the CPL parametrisation \cite{Chevallier:2001,Linder:2003}
for the dark-energy equation of state,
\begin{equation}
  w(a) = w_0 + w_a\,(1 - a)\,,
  \label{eq:wCPL}
\end{equation}
where $a = (1+z)^{-1}$.  In a spatially flat universe with matter,
radiation, and dark energy, the normalised Hubble rate is
\begin{equation}
  E^2(z) = \Omega_m(1+z)^3 + \Omega_r(1+z)^4
  + \Omega_{\rm DE}\,(1+z)^{3(1+w_0+w_a)}\,
    \exp\!\left[-\frac{3\,w_a\,z}{1+z}\right],
  \label{eq:Ez}
\end{equation}
with $\Omega_{\rm DE} = 1 - \Omega_m - \Omega_r$ and
$\Omega_r = \Omega_m/(1 + z_{\rm eq})$, $z_{\rm eq} = 3387$
\cite{Planck:2020}.  The three distance measures relevant for BAO are
the comoving distance
\begin{equation}
  \DM(z) = \frac{c}{H_0}\int_0^z \frac{\mathrm{d}z'}{E(z')}\,,
  \label{eq:DM}
\end{equation}
the Hubble distance $\DH(z) = c/[H_0\,E(z)]$, and the
volume-averaged distance
$\DV(z) = [z\,\DM(z)^2\,\DH(z)]^{1/3}$.  All are reported in units
of $\rd$, so the dependence on $H_0$ partially cancels but does not
vanish: $\DM/\rd = (c/H_0\rd)\int_0^z \mathrm{d}z'/E(z')$ depends
on $H_0$ through the prefactor $c/(H_0\rd)$.

\subsubsection*{CMB-consistent $(\Omega_m, H_0)$ at each $(w_0, w_a)$}

We fix $\rd = 147.09$\,Mpc \cite{Planck:2020} and determine both
$\Omega_m$ and $H_0$ from two constraints: the Planck angular scale
$\theta_*$ and the best fit to the DESI data.

The primary CMB geometric observable is
$\theta_* = r_*/\DM(z_*)$, measured with fractional precision
$0.030\%$ \cite{Planck:2020}.  Since CPL modifies only the
post-recombination expansion, $r_*$ is unchanged and the constraint
fixes $\DM(z_*)$.  Writing
$\DM(z_*) = (c/H_0)\int_0^{z_*}\mathrm{d}z'/E(z';\Omega_m,w_0,w_a)$,
the $\theta_*$ constraint determines $H_0$ as a function of
$\Omega_m$ at each $(w_0, w_a)$:
\begin{equation}
  H_0(\Omega_m, w_0, w_a)
  = H_0^{\rm Planck}\;\frac{\displaystyle\int_0^{z_*}
    \mathrm{d}z'/E(z';\Omega_m, w_0, w_a)}
    {\displaystyle\int_0^{z_*}
    \mathrm{d}z'/E(z';\Omega_m^{\rm Planck}, -1, 0)}\,,
  \label{eq:H0_theta}
\end{equation}
where $H_0^{\rm Planck} = 67.36$\,km\,s$^{-1}$\,Mpc$^{-1}$ and
$\Omega_m^{\rm Planck} = 0.3153$.  This ensures $\DM(z_*) =
\DM^{\LCDM}(z_*)$ by construction, so that $\theta_*$ matches the
Planck measurement.

With $H_0(\Omega_m)$ fixed by Eq.~\eqref{eq:H0_theta}, the
remaining free parameter $\Omega_m$ is determined by minimising the
DESI $\chi^2$:
\begin{equation}
  \Omega_m^{\rm best}(w_0, w_a)
  = \underset{\Omega_m}{\arg\min}\;
    \chi^2_{\rm DESI}(\Omega_m, H_0(\Omega_m), w_0, w_a)\,.
  \label{eq:Om_desi}
\end{equation}
This is a one-dimensional minimisation at each grid point, solved
numerically over $0.15 < \Omega_m < 0.55$.  The procedure yields, at
each $(w_0, w_a)$, the CMB-consistent CPL model that \emph{best
represents the DESI data}, enabling a fair comparison with BAOtr: we
ask whether the model that optimally describes DESI also describes
BAOtr.

We note that matching $\theta_*$ alone does not guarantee consistency
with the full CMB power spectrum; models with extreme $(w_0, w_a)$
may require $\Omega_m$ or $H_0$ values that conflict with the CMB
peak structure.  A full MCMC with Planck likelihoods would be needed
to assess this, but the geometric constraint captures the dominant
degeneracy.\footnote{The small difference between our
$\Omega_m^{\rm best}$ and the marginalised values of
Xu~et~al.~\cite{Xu:2026} arises because their MCMC also adjusts
$\Omega_b h^2$ and $n_s$ to fit the full CMB spectrum, not just
$\theta_*$.}

Table~\ref{tab:Om_values} lists the DESI-optimized parameters at the
published CPL posterior centres.  The CMB+DESI model recovers
$\Omega_m = 0.352$, $H_0 = 63.8$\,km\,s$^{-1}$\,Mpc$^{-1}$ ---
consistent with the Xu~et~al.\ posterior ($\Omega_m \approx 0.350$,
$H_0 \approx 63.9$), confirming that our procedure captures the
correct parameter degeneracy.

\begin{table}[ht]
\centering
\caption{DESI-optimized parameters at each $(w_0, w_a)$: $\Omega_m$
minimises $\chi^2_{\rm DESI}$ with $H_0$ set by the $\theta_*$
constraint (Eq.~\ref{eq:H0_theta}).  Our values are deterministic at
fixed $(w_0, w_a)$; the Xu~et~al.\ column shows marginalised
posteriors from full MCMC fits \cite{Xu:2026}.  The two differ because
the MCMC also adjusts $\Omega_b h^2$ and $n_s$.}
\label{tab:Om_values}
\setlength{\tabcolsep}{3pt}
\footnotesize
\resizebox{\columnwidth}{!}{%
\begin{tabular}{@{}lrrcccc@{}}
\toprule
Model & $w_0$ & $w_a$ & $\Omega_m$ & $H_0$
  & $\Omega_m^{\rm Xu}$ & $H_0^{\rm Xu}$ \\
  & & & \multicolumn{2}{c}{(this work)}
  & \multicolumn{2}{c}{(Xu et al.)} \\
\midrule
$\LCDM$
  & $-1.000$ & $\phantom{-}0.000$
  & 0.299 & 68.8
  & $0.3153\!\pm\!0.0073$ & $67.36\!\pm\!0.54$ \\
CMB+PP\&SH0ES
  & $-0.694$ & $-1.700$
  & 0.318 & 67.7
  & $0.283\!\pm\!0.006$ & $70.87\!\pm\!0.68^a$ \\
CMB+PP\&SH0ES+BAOtr
  & $-0.660$ & $-1.910$
  & 0.319 & 67.6
  & $0.279\!\pm\!0.006$ & $71.31\!\pm\!0.67^a$ \\
CMB+SDSS
  & $-0.480$ & $-1.510$
  & 0.347 & 64.1
  & $0.355\!\pm\!0.026$ & $63.6^{+2.2}_{-2.5}$ \\
CMB+DESI
  & $-0.420$ & $-1.750$
  & 0.352 & 63.8
  & $0.350\!\pm\!0.021$ & $63.9\!\pm\!2.0$ \\
\bottomrule
\end{tabular}%
}
\par\vspace{2pt}
\begin{minipage}{\columnwidth}
\footnotesize $^a$\,For these two combinations the Xu~et~al.\ $H_0$
posterior is dominated by the SH0ES Cepheid calibration
incorporated into the SN~Ia likelihood, which acts as a near-local
$H_0$ constraint.  Our $H_0$ is derived from the Planck $\theta_*$
constraint alone and targets a different quantity.
\end{minipage}
\end{table}

\subsubsection*{Parameter space and prior}

We scan
\begin{equation}
  w_0 \in [-2.5,\;0.5]\,,\qquad
  w_a \in [-5.5,\;2.5]\,,
  \label{eq:scan_range}
\end{equation}
subject to $w_0 + w_a < 0$ (early-matter-domination prior).  At each
grid point, $\Omega_m$ and $H_0$ are determined by
Eqs.~\eqref{eq:H0_theta}--\eqref{eq:Om_desi}, so the scan explores
the one-dimensional DESI-optimized, CMB-consistent subspace at each
$(w_0, w_a)$.

\subsection{Tension statistic}
\label{sec:statistic}

We quantify the BAOtr--BAO\,3D tension using two complementary
methods.  Method~A (direct model comparison) serves as the primary
analysis; Method~B ($\alpha$-interpolation) provides an independent
cross-check.

\subsubsection*{Method A: direct model comparison (primary)}

At each $(w_0, w_a)$ with DESI-optimized $(\Omega_m, H_0)$, we
compute the model predictions and compare directly with both
datasets:
\begin{equation}
  \chi^2_{\rm BAOtr}(w_0, w_a)
  = \sum_{i=1}^{15}
    \left[\frac{(\DM/\rd)_i^{\rm BAOtr}
                - (\DM/\rd)^{\mathcal{M}}(z_i)}
               {\sigma_{{\rm BAOtr},i}}\right]^2,
  \label{eq:chi2_baotr}
\end{equation}
\begin{equation}
  \chi^2_{\rm DESI}(w_0, w_a)
  = \sum_{j=1}^{N_{\rm anc}}
    \left[\frac{X_j^{\rm data}
                - X_j^{\mathcal{M}}(z_j)}
               {\sigma_j}\right]^2,
  \label{eq:chi2_desi}
\end{equation}
In Eq.~\eqref{eq:chi2_desi} $X_j$ denotes $\DM/\rd$ for the six anisotropic tracers and $\DV/\rd$ for BGS.  $\chi^2_{\rm DESI}$ is minimised at each
$(w_0, w_a)$; the question is whether the DESI-optimized model also
fits BAOtr.  The 15 BAOtr points are treated as independent (disjoint
shells); we note that shared SDSS angular systematics could introduce
correlated biases, effectively reducing the number of independent
degrees of freedom (Section~\ref{sec:limitations}).

We scan on a $55 \times 55$ grid over the range of
Eq.~\eqref{eq:scan_range}.

\subsubsection*{Method B: $\alpha$-interpolation (cross-check)}

Method~A asks which model best fits each dataset.  Method~B asks a
complementary question: \emph{accepting the DESI data as fixed
anchors}, does CPL provide a better interpolation between them than
$\LCDM$?

At each DESI anchor redshift $z_j$, we define
\begin{equation}
  \alpha_j(w_0, w_a)
  = \frac{(\DM/\rd)_j^{\rm data}}{(\DM/\rd)^{\mathcal{M}}(z_j)}\,,
  \label{eq:alpha_def}
\end{equation}
using the DESI-optimized $(\Omega_m, H_0)$.  For the BGS anchor,
$(\DM/\rd)^{\rm data}$ is obtained from the measured $\DV/\rd$ using
the model-predicted $\DH/\rd$:
$\DM/\rd = [(\DV/\rd)^3/(z\cdot\DH^{\mathcal{M}}/\rd)]^{1/2}$.
A continuous $\alpha(z)$ is constructed by piecewise linear
interpolation of $\{z_j, \alpha_j\}$ in $u = \ln(1+z)$, with
constant-$\alpha$ extrapolation below the lowest anchor (alternatives
tested in Section~\ref{sec:robustness}).  The predicted $\DM/\rd$ at
each BAOtr redshift is
\begin{equation}
  \left(\frac{\DM}{\rd}\right)_i^{\!\rm pred}
  = \alpha(z_i) \times
    \left(\frac{\DM}{\rd}\right)^{\!\mathcal{M}}\!(z_i)\,,
  \label{eq:pred_DM}
\end{equation}
and the tension statistic is
\begin{equation}
  T_i = \frac{(\DM/\rd)_i^{\rm BAOtr}
          - (\DM/\rd)_i^{\rm pred}}
         {\sqrt{\sigma_{{\rm BAOtr},i}^2
                + \sigma_{{\rm pred},i}^2}}\,,
  \qquad
  \chi^2_B = \sum_{i=1}^{15} T_i^2\,,
  \label{eq:Ti}
\end{equation}
where $\sigma_{{\rm pred},i}$ propagates the DESI uncertainties
through the interpolation (subdominant at all redshifts).

The key distinction: Method~A allows the model to miss the DESI
points and measures the absolute fit to each dataset.  Method~B
forces the prediction through the DESI data (via $\alpha$-warping)
and tests the \emph{shape} between anchors.  As we show in
Section~\ref{sec:consistency}, both reach the same conclusion.

%% ═══════════════════════════════════════════════════════════════════
\section{Results}
\label{sec:results}
%% ═══════════════════════════════════════════════════════════════════

\subsection{Baseline tension under \texorpdfstring{$\LCDM$}{LCDM}}
\label{sec:baseline}

Table~\ref{tab:baseline} presents the per-point BAOtr tension under
the DESI-optimized $\LCDM$ model ($\Omega_m = 0.299$,
$H_0 = 68.8$\,km\,s$^{-1}$\,Mpc$^{-1}$, $\chi^2_{\rm DESI} = 5.4$)
for both methods.

\begin{table}[ht]
\centering
\caption{Per-point BAOtr tension under the DESI-optimized $\LCDM$
model ($\Omega_m = 0.299$, $H_0 = 68.8$\,km\,s$^{-1}$\,Mpc$^{-1}$).
$T_i^A$: Method~A (direct model); $T_i^B$: Method~B
($\alpha$-interpolation with BGS anchor, constant-$\alpha$
extrapolation).  Negative values indicate BAOtr prefers smaller
$\DM/\rd$.}
\label{tab:baseline}
\setlength{\tabcolsep}{4.5pt}
\footnotesize
\begin{tabular}{clcccccc}
\toprule
& & & & \multicolumn{2}{c}{Method A (direct)}
      & \multicolumn{2}{c}{Method B ($\alpha$)} \\
\cmidrule(lr){5-6} \cmidrule(lr){7-8}
\# & $z$ & $(\DM/\rd)^{\rm BAOtr}$ & $\sigma$
   & $T_i^A\;[\sigma]$ & $(\chi^2_i)^A$
   & $T_i^B\;[\sigma]$ & $(\chi^2_i)^B$ \\
\midrule
 1 & 0.110 &  2.894 & 0.477 & $-0.59$ &  0.35 & $-0.61$ & 0.38 \\
 2 & 0.235 &  6.324 & 0.161 & $-1.58$ &  2.49 & $-1.50$ & 2.25 \\
 3 & 0.365 &  9.052 & 0.315 & $-2.64$ &  6.95 & $-2.82$ & 7.93 \\
 4 & 0.450 & 12.012 & 0.428 & $+0.23$ &  0.05 & $-0.25$ & 0.06 \\
 5 & 0.470 & 11.414 & 0.568 & $-1.70$ &  2.88 & $-2.06$ & 4.25 \\
 6 & 0.490 & 11.483 & 0.483 & $-2.80$ &  7.82 & $-3.21$ & 10.30 \\
 7 & 0.510 & 11.912 & 0.421 & $-3.27$ & 10.71 & $-3.70$ & 13.70 \\
 8 & 0.530 & 13.356 & 0.934 & $-0.41$ &  0.17 & $-0.68$ & 0.47 \\
 9 & 0.550 & 13.481 & 0.793 & $-0.88$ &  0.78 & $-1.15$ & 1.33 \\
10 & 0.570 & 12.402 & 1.074 & $-2.06$ &  4.23 & $-2.23$ & 4.96 \\
11 & 0.590 & 13.111 & 1.050 & $-1.84$ &  3.39 & $-1.98$ & 3.92 \\
12 & 0.610 & 14.843 & 1.269 & $-0.50$ &  0.25 & $-0.59$ & 0.34 \\
13 & 0.630 & 14.767 & 1.598 & $-0.71$ &  0.50 & $-0.75$ & 0.56 \\
14 & 0.650 & 16.185 & 0.777 & $-0.16$ &  0.03 & $-0.20$ & 0.04 \\
15 & 2.225 & 32.371 & 5.667 & $-0.99$ &  0.99 & $-1.01$ & 1.02 \\
\midrule
\multicolumn{4}{r}{\textbf{Total $\chi^2$}}
   & & $\mathbf{41.6}$ & & $\mathbf{51.5}$ \\
\bottomrule
\end{tabular}
\end{table}

Under Method~A, the global tension is $\chi^2_{\rm BAOtr} = 41.6$ for
15 points ($p \approx 2 \times 10^{-4}$, $3.5\sigma$), while the
DESI fit is good ($\chi^2_{\rm DESI} = 5.4$ for 7 points).  Under
Method~B, the DESI-anchored prediction gives $\chi^2_B = 51.5$,
higher because the $\alpha$-warping amplifies the model--data offset
at the anchor redshifts.

In Method A, all but one of the 15 tensions are negative; in Method B, all 15 are negative (BAOtr prefers smaller $\DM/\rd$), and the tension is
concentrated at $z = 0.35$--$0.60$.  The five points at $z = 0.365$,
$0.470$, $0.490$, $0.510$, and $0.570$ contribute
$\chi^2 \approx 32$--$35$ depending on method, accounting for
$\sim$\!78--82\% of the total.  The most significant point is
$z = 0.510$, where the DESI-optimized $\LCDM$ prediction is
$\DM/\rd = 13.29$ while BAOtr gives $11.91 \pm 0.42$, a $3.3\sigma$
discrepancy contributing $\chi^2 \approx 10.7$.  This per-point
$\chi^2$ contribution sets a robust floor on the BAOtr--DESI
tension.  The model-independent data-versus-data comparison at this
redshift (Section~\ref{sec:baotr_data}) yields an even larger
$3.7\sigma$ clash, confirming that no smooth modification of
$\DM(z)$ can bring the two datasets into agreement.

\begin{figure}[htbp]
\centering
  \includegraphics[width=0.90\textwidth]{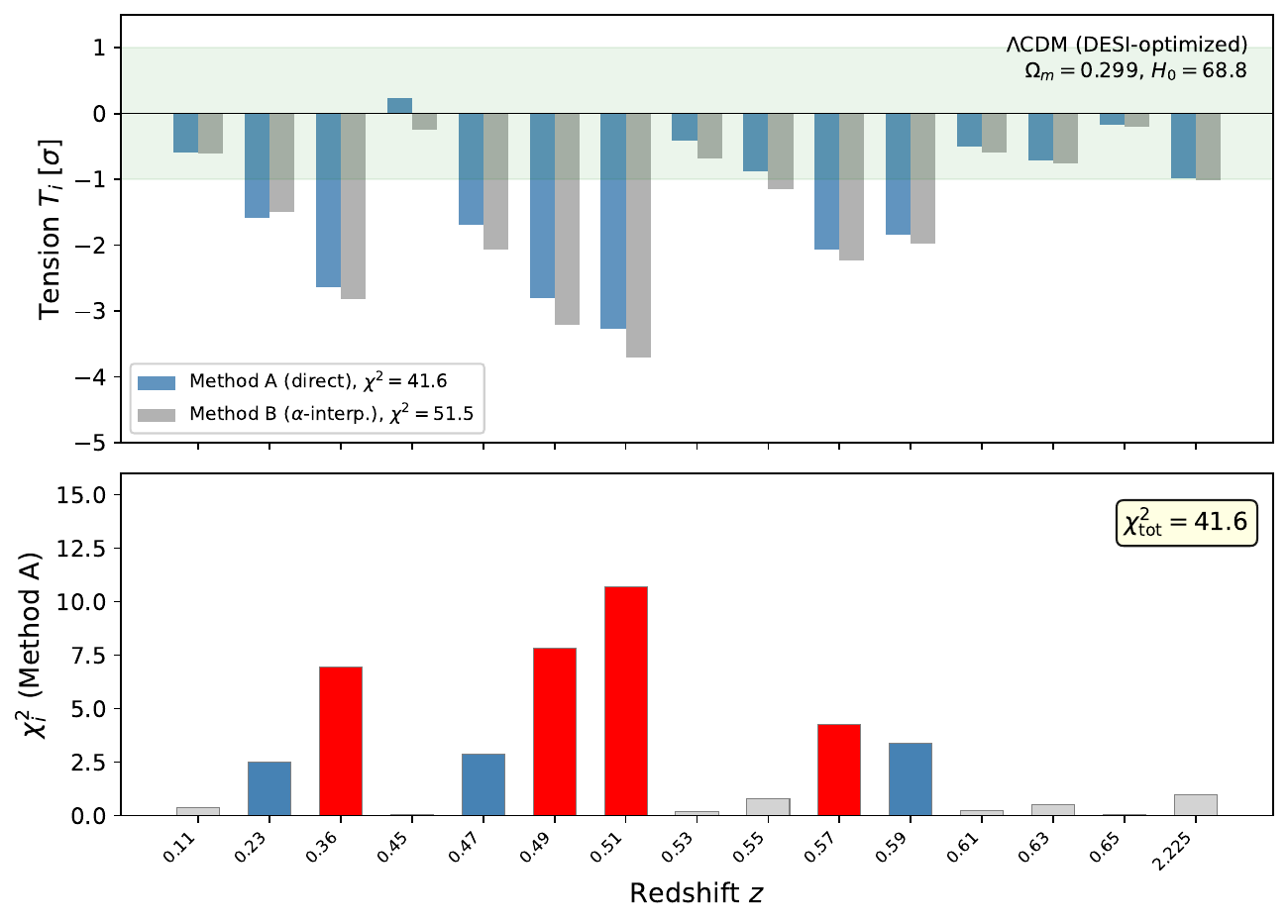}
  \caption{Per-redshift BAOtr tension under DESI-optimized $\LCDM$
  ($\Omega_m = 0.299$, $H_0 = 68.8$).
  \textbf{Upper:} normalised tension $T_i$; blue bars show
  Method~A, grey bars Method~B.  The green band marks
  $|T_i| < 1\sigma$.
  \textbf{Lower:} per-point $\chi^2_i$ (Method~A).
  Red: $\chi^2_i > 4$; blue: $1 < \chi^2_i \leq 4$;
  grey: $\chi^2_i \leq 1$.}
  \label{fig:tension_bars}
\end{figure}

Figure~\ref{fig:tension_bars} displays the per-point tensions
graphically.

\subsection{No CMB-consistent CPL model resolves the tension}
\label{sec:cpl_results}

\subsubsection*{The DESI--BAOtr trade-off}

\begin{figure}[htbp]
\centering
  \includegraphics[width=0.80\textwidth]{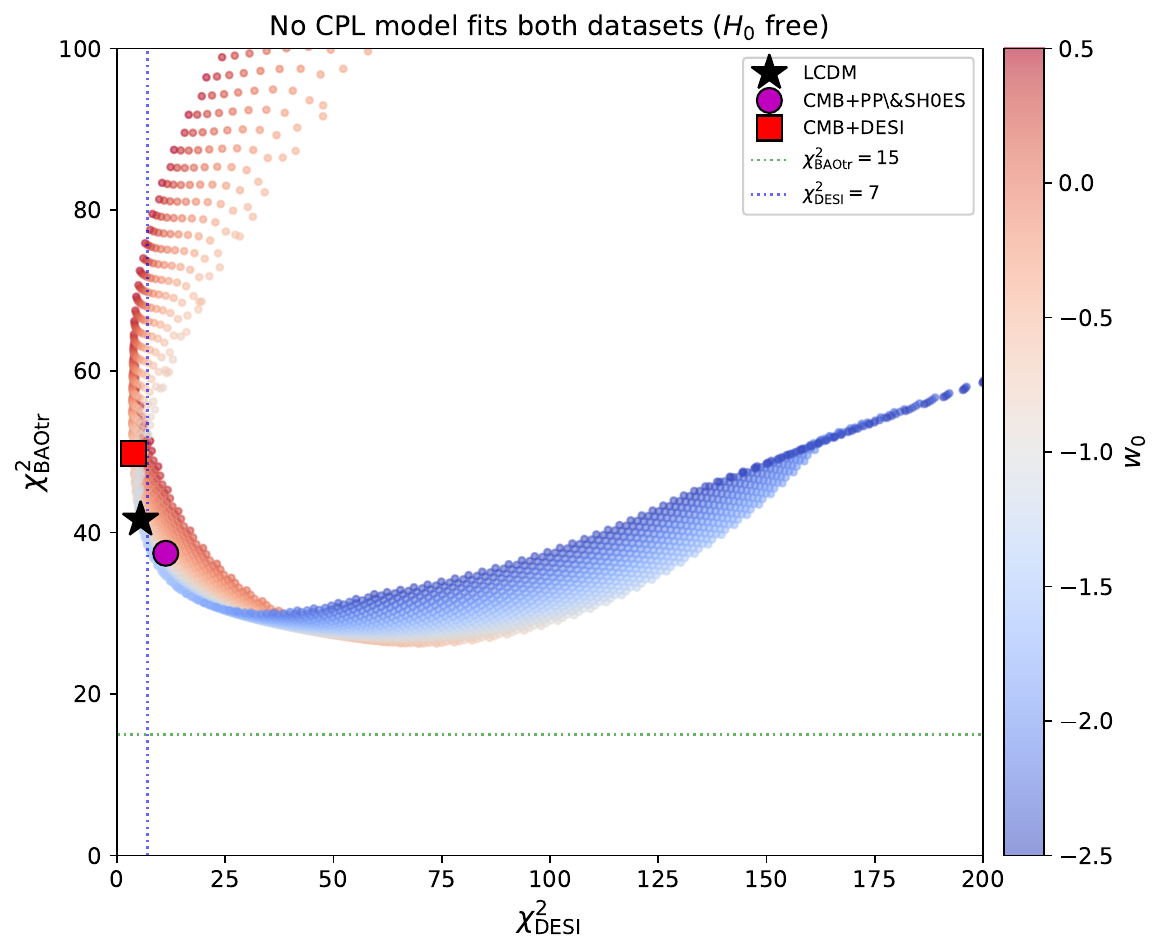}
  \caption{Trade-off between $\chi^2_{\rm DESI}$ and
  $\chi^2_{\rm BAOtr}$ (Method~A) across the DESI-optimized,
  CMB-consistent CPL parameter space.  Each point represents a
  $(w_0, w_a)$ model with $(\Omega_m, H_0)$ set by
  Eqs.~\eqref{eq:H0_theta}--\eqref{eq:Om_desi}, coloured by $w_0$.
  The black star marks $\LCDM$; the magenta circle CMB+PP\&SH0ES;
  the red square CMB+DESI.  Dotted lines show the expected $\chi^2$
  for consistency.  No CPL model reaches the lower-left corner.}
  \label{fig:tradeoff}
\end{figure}

Figure~\ref{fig:tradeoff} shows $\chi^2_{\rm DESI}$ vs
$\chi^2_{\rm BAOtr}$ (Method~A) for each DESI-optimized model in the
scan.  The anti-correlation is stark: no model falls in the
lower-left corner. Among all models with $\chi^2_{\rm DESI} < 20$
(622 models), the minimum $\chi^2_{\rm BAOtr}$ is 31.6.  Among all
models with $\chi^2_{\rm DESI} < 15$, the minimum $\chi^2_{\rm BAOtr}$
is 32.9.  There are zero models with $\chi^2_{\rm DESI} < 20$ and
$\chi^2_{\rm BAOtr} < 25$ simultaneously.

\subsubsection*{$\chi^2$ summary}

Table~\ref{tab:chi2_summary} presents the $\chi^2$ values for both
methods at the published CPL posteriors.

\begin{table}[ht]
\centering
\caption{Tension $\chi^2$ for DESI-optimized CPL models under both
methods.  $\Delta\chi^2 = \chi^2 - \chi^2_{\LCDM}$: positive values
mean the BAOtr tension increases.  Method~A reports
$\chi^2_{\rm BAOtr}$ and $\chi^2_{\rm DESI}$; Method~B reports the
DESI-anchored $\chi^2_B$.}
\label{tab:chi2_summary}
\setlength{\tabcolsep}{3.5pt}
\footnotesize
\begin{tabular}{lcccccccc}
\toprule
& & & & & \multicolumn{2}{c}{Method A}
      & \multicolumn{2}{c}{Method B} \\
\cmidrule(lr){6-7} \cmidrule(lr){8-9}
Model & $w_0$ & $w_a$ & $\Omega_m$ & $H_0$
  & $\chi^2_{\rm BAOtr}$ & $\chi^2_{\rm DESI}$
  & $\chi^2_B$ & $\Delta\chi^2_B$ \\
\midrule
$\LCDM$
  & $-1.00$ & $0.00$ & 0.299 & 68.8
  & 41.6 & 5.4
  & 51.5 & --- \\[3pt]
\multicolumn{9}{l}{\textit{Published posteriors \cite{Xu:2026}:}} \\
CMB+PP\&SH0ES
  & $-0.69$ & $-1.70$ & 0.318 & 67.7
  & 37.4 & 11.2
  & 53.0 & $+1.5$ \\
CMB+PP\&SH0ES+BAOtr
  & $-0.66$ & $-1.91$ & 0.319 & 67.6
  & 36.9 & 12.5
  & 53.2 & $+1.6$ \\
CMB+SDSS
  & $-0.48$ & $-1.51$ & 0.347 & 64.1
  & 49.7 & 3.7
  & 54.3 & $+2.8$ \\
CMB+DESI
  & $-0.42$ & $-1.75$ & 0.352 & 63.8
  & 49.8 & 3.7
  & 54.7 & $+3.2$ \\
\bottomrule
\end{tabular}
\end{table}

The two methods tell a consistent story:

\begin{enumerate}
\item \textbf{Method~A:} The CMB+DESI CPL model fits DESI well
  ($\chi^2_{\rm DESI} = 3.7$) but makes the BAOtr tension
  \emph{worse} ($\chi^2_{\rm BAOtr} = 49.8$, up from 41.6 under
  $\LCDM$).  This model has $\Omega_m = 0.352$ and
  $H_0 = 63.8$\,km\,s$^{-1}$\,Mpc$^{-1}$: the lower $H_0$
  increases $\DM/\rd$ (since $\DM \propto c/H_0$), pushing the
  prediction above BAOtr.  Conversely, the CMB+PP\&SH0ES model
  modestly improves the BAOtr fit ($\chi^2_{\rm BAOtr} = 37.4$) but
  worsens the DESI fit ($\chi^2_{\rm DESI} = 11.2$).

\item \textbf{Method~B:} All published CPL posteriors \emph{increase}
  the DESI-anchored tension ($\Delta\chi^2_B = +1.5$ to $+3.2$).
  This occurs because CPL with $w_0 > -1$ predicts smaller $\DM/\rd$
  than $\LCDM$ at the anchor redshifts (at matched $H_0$), inflating
  $\alpha_j$ above unity and pushing the interpolated prediction
  further from BAOtr.
\end{enumerate}

\subsubsection*{$\chi^2$ surfaces}

\begin{figure}[htbp]
\centering
  \includegraphics[width=\textwidth]{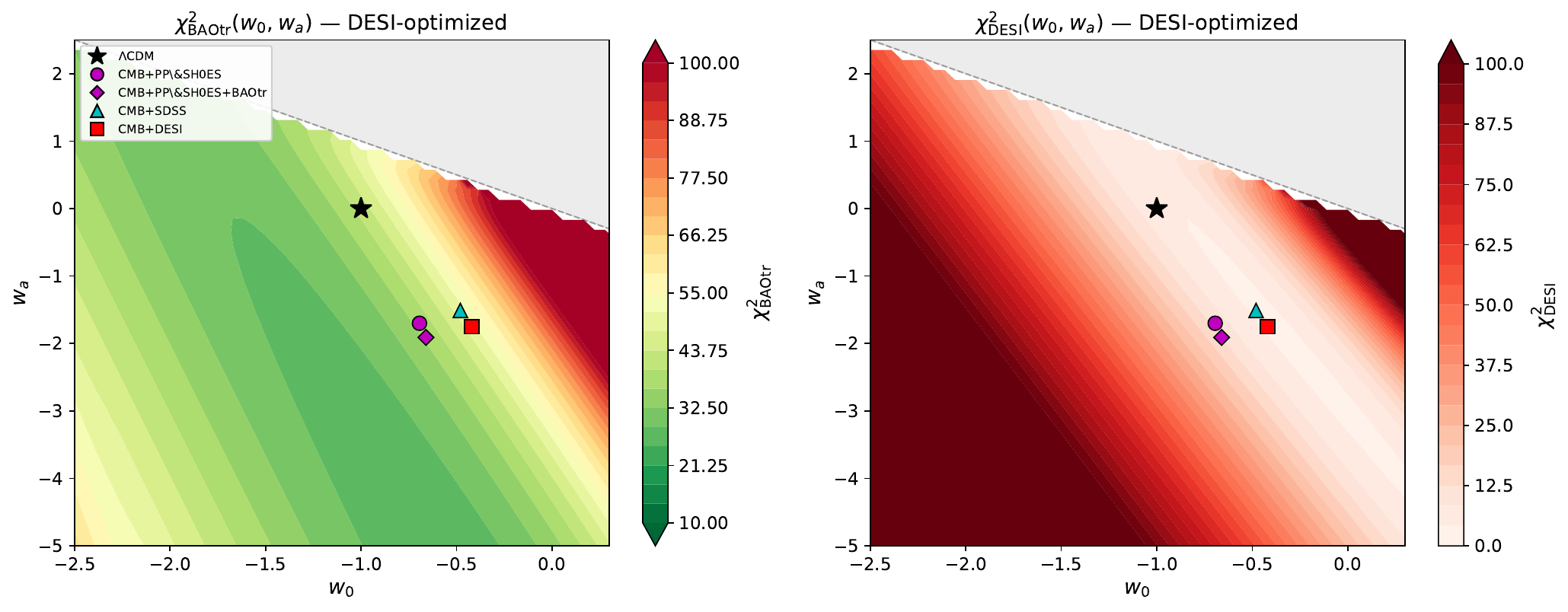}
  \caption{\textbf{Left:} $\chi^2_{\rm BAOtr}(w_0, w_a)$ (Method~A,
  DESI-optimized).  Green: low tension.
  \textbf{Right:} $\chi^2_{\rm DESI}(w_0, w_a)$.  Light: low
  $\chi^2$.  The two surfaces have opposite gradients, confirming the
  trade-off.  Symbols mark published posteriors as in
  Fig.~\ref{fig:tradeoff}.}
  \label{fig:chi2_surface}
\end{figure}

Figure~\ref{fig:chi2_surface} shows the $\chi^2_{\rm BAOtr}$ and
$\chi^2_{\rm DESI}$ surfaces across the $(w_0, w_a)$ plane, each
DESI-optimized.  The two surfaces have opposite gradients: moving
toward better BAOtr agreement (upper-left, more negative $w_0$)
worsens the DESI fit, and vice versa.

\subsubsection*{Why CPL cannot resolve the tension}

The failure has a transparent origin.  BAOtr prefers smaller $\DM/\rd$
than DESI at $z \lesssim 0.65$.  To fit BAOtr, the model needs
smaller $\DM/\rd$ at these redshifts, achieved by increasing $H_0$
(since $\DM \propto c/H_0$).  But the $\theta_*$ constraint links
$H_0$ to $\Omega_m$: increasing $H_0$ requires decreasing $\Omega_m$,
which changes $E(z)$ and worsens the DESI fit.  The direct clash at
$z = 0.510$ ($3.7\sigma$, data-versus-data) is the most acute manifestation: both
datasets measure the same physical distance at the same redshift, so
no modification of $\DM(z)$ can reconcile them.

\subsection{Robustness tests}
\label{sec:robustness}

\subsubsection*{Extrapolation and BGS anchor (Method~B)}

Table~\ref{tab:sensitivity} presents the $\LCDM$ baseline $\chi^2_B$
for six configurations of Method~B.

\begin{table}[ht]
\centering
\caption{Method~B $\LCDM$ baseline tension ($\chi^2_B$ for 15 BAOtr
points) under three extrapolation schemes, all including the BGS
anchor at $z = 0.295$.  $\chi^2_B/N \geq 3.3$ in all cases.}
\label{tab:sensitivity}
\setlength{\tabcolsep}{6pt}
\begin{tabular}{lcc}
\toprule
Extrapolation & $\chi^2_B$ & $\chi^2_B/N_{\rm pts}$ \\
\midrule
Constant-$\alpha$ (default) & 51.5 & 3.43 \\
Model ($\alpha = 1$)        & 51.1 & 3.41 \\
Linear                      & 50.8 & 3.39 \\
\bottomrule
\end{tabular}
\end{table}

The three extrapolation schemes give $\chi^2_B = 50.8$--$51.5$
(spread $\Delta\chi^2 = 0.7$), much smaller than the baseline tension
itself.  The dominant contributions ($z = 0.35$--$0.65$) lie within
the interpolation range and are unaffected by the extrapolation
choice.

\subsubsection*{SDSS vs DESI}

\begin{figure}[htbp]
\centering
  \includegraphics[width=0.85\textwidth]{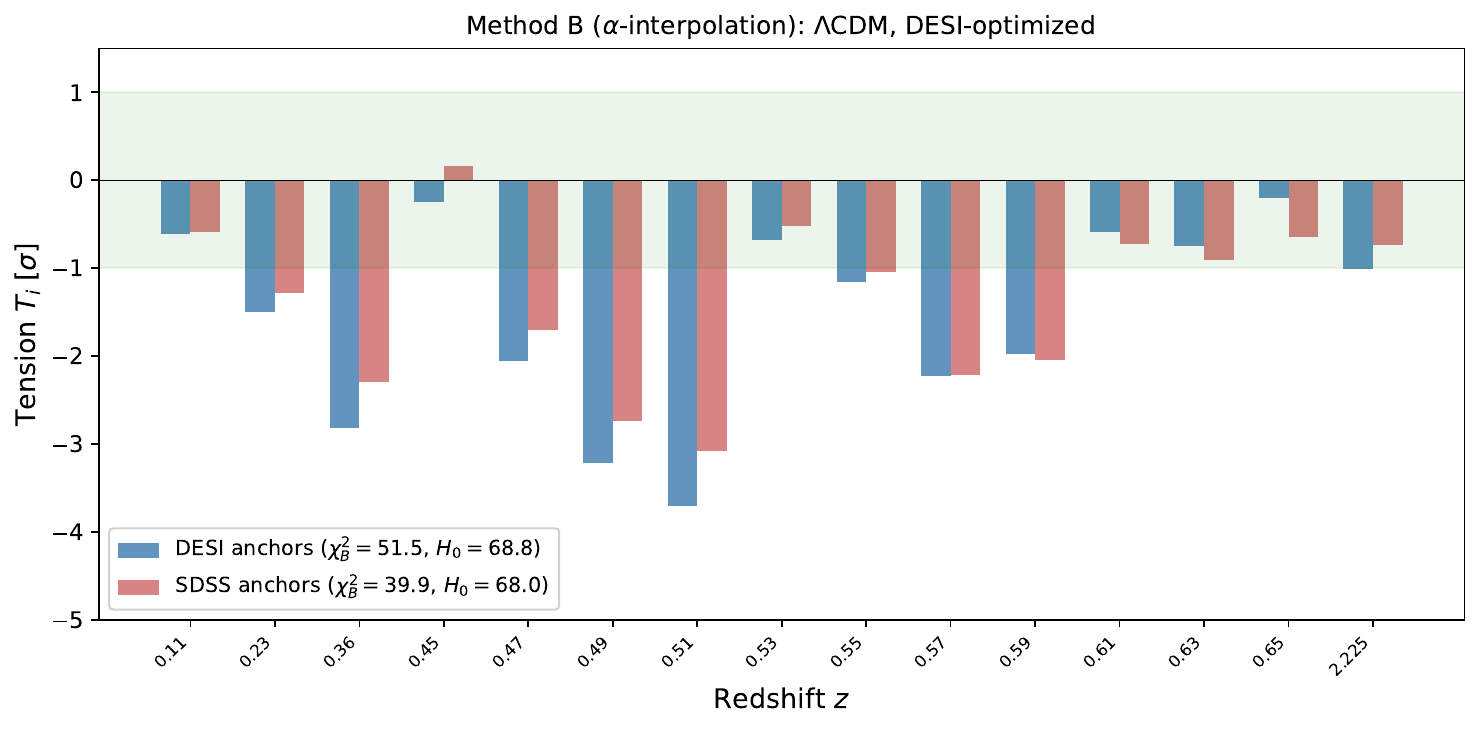}
  \caption{Per-point tension $T_i$ (Method~B, $\LCDM$) using DESI
  anchors (blue) and SDSS anchors (red).  Both produce the same
  pattern.  The tension is a generic feature of the BAOtr--BAO\,3D
  comparison.}
  \label{fig:sdss_comparison}
\end{figure}

Under Method~B with SDSS anchors, the $\LCDM$ tension is
$\chi^2_B = 39.9$ ($H_0 = 68.0$\,km\,s$^{-1}$\,Mpc$^{-1}$),
somewhat lower than the DESI value of 51.5 but still highly
significant ($\chi^2/N = 2.7$, $3.3\sigma$).  The per-point pattern
is strikingly similar (Fig.~\ref{fig:sdss_comparison}): all tensions
negative, concentrated at $z = 0.35$--$0.60$, dominated by the same
redshift bins.  The persistence across two independent surveys rules
out DESI-specific artefacts.

\subsection{Consistency between methods}
\label{sec:consistency}

\begin{figure}[htbp]
\centering
  \includegraphics[width=\textwidth]{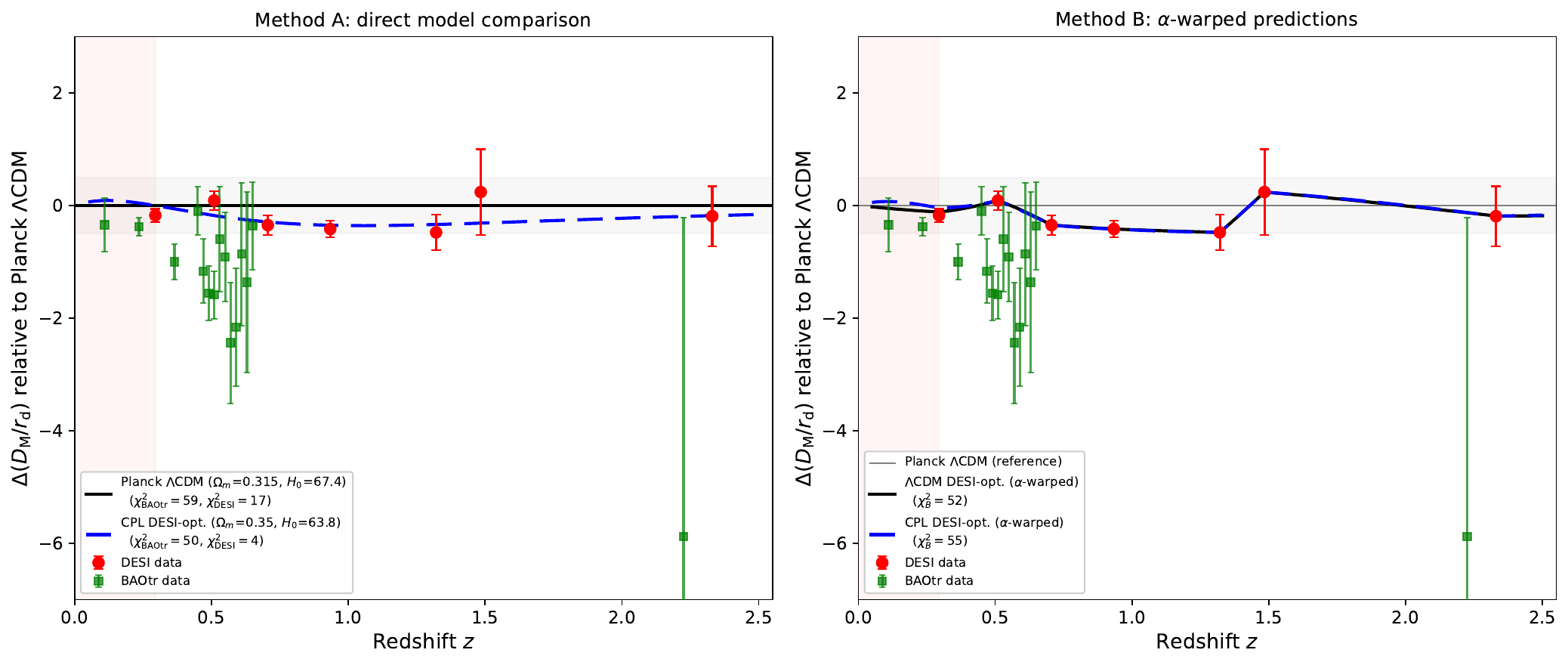}
  \caption{Residuals $\Delta(\DM/\rd)$ relative to Planck $\LCDM$
  ($\Omega_m = 0.3153$, $H_0 = 67.36$\,km\,s$^{-1}$\,Mpc$^{-1}$).
  DESI data (red circles) and BAOtr data (green squares) are shown
  with their $1\sigma$ uncertainties.  The pink shaded region
  ($z < 0.295$) marks the zone below the lowest DESI anchor,
  where Method~B predictions depend on the extrapolation scheme.
  \textbf{Left (Method~A):} Planck $\LCDM$ defines the zero line
  (black solid; $\chi^2_{\rm BAOtr} = 59$,
  $\chi^2_{\rm DESI} = 17$).  The DESI-optimized CPL model
  ($\Omega_m = 0.35$, $H_0 = 63.8$; blue dashed) sits slightly
  below Planck $\LCDM$ at $z \lesssim 1$ but cannot bridge the gap
  to BAOtr ($\chi^2_{\rm BAOtr} = 50$, $\chi^2_{\rm DESI} = 4$).
  \textbf{Right (Method~B):} $\alpha$-warped predictions using
  DESI-optimized parameters, shown as residuals relative to Planck
  $\LCDM$.  The grey horizontal line marks the Planck $\LCDM$
  reference.  The DESI-optimized $\LCDM$ prediction (black solid;
  $\chi^2_B = 52$) and the CPL prediction (blue dashed;
  $\chi^2_B = 55$) are both pinned near zero at the DESI anchor
  redshifts by construction, but BAOtr remains 1--3 units below.
  In both panels, no model approaches the BAOtr data at
  $z = 0.35$--$0.65$.}
  \label{fig:methods_comparison}
\end{figure}

Figure~\ref{fig:methods_comparison} illustrates why the two methods
give different $\chi^2$ values but the same conclusion.

Under Method~A (left panel), the CMB+DESI CPL model passes through
the DESI data (as intended by the DESI optimization), but its lower
$H_0 = 63.8$ increases all distances, placing the curve above BAOtr.
Under Method~B (right panel), the $\alpha$-warping forces both models
through DESI, but CPL's smaller $(\DM/\rd)^{\mathcal{M}}$ at the
anchors inflates $\alpha_j > 1$, pushing the interpolated prediction
above $\LCDM$.

Both perspectives confirm the same conclusion:
\begin{itemize}
\item Method~A reveals the tension as a \emph{dataset incompatibility}:
  no CPL model achieves low $\chi^2$ for both DESI and BAOtr
  (Fig.~\ref{fig:tradeoff}).
\item Method~B reveals the tension as an \emph{interpolation failure}:
  anchoring to DESI and varying the CPL backbone cannot bring the
  predictions into agreement with BAOtr.
\end{itemize}

%% ═══════════════════════════════════════════════════════════════════
\section{Discussion}
\label{sec:discussion}
%% ═══════════════════════════════════════════════════════════════════

\subsection{Fiducial independence and the rescaling error}
\label{sec:discuss_fiducial}

The fiducial independence of published 3D BAO distances
(Section~\ref{sec:pipeline}) is a straightforward algebraic identity:
$\alpha_\perp$ is defined as the ratio of true to fiducial distances
(Eq.~\ref{eq:alpha_solved}), and the pipeline publishes
$\alpha_\perp \times \DM^{\rm fid}/\rd^{\rm fid}$, in which the
fiducial cancels by construction (Eq.~\ref{eq:cancellation}).
Residual effects are $\delta\alpha/\alpha \lesssim 0.1$--$0.3\%$
\cite{AbdulKarim:2025,PerezFernandez:2024}, two orders of magnitude
below the BAOtr uncertainties.  The rescaling approach overcorrects by
a factor of 20--50 (Table~\ref{tab:overcorrection}), producing a
spurious $\Delta\chi^2 \sim 40$--$50$.

\subsection{Remaining explanations for the tension}
\label{sec:explanations}

With fiducial misspecification ruled out and CPL unable to reconcile
the datasets (Section~\ref{sec:cpl_results}), two explanations remain.
A breakdown of the distance-duality relation has been tested and
ruled out by Favale~et~al.~\cite{Favale:2024} for both 2D and
3D BAO under the assumption of a transparent intergalactic medium;
a smooth cosmic opacity could redistribute the apparent tension
across redshifts but cannot erase the direct $z = 0.510$ clash,
which is insensitive to any such factor.

\subsubsection*{BAOtr systematics}

Projection effects from finite shell width can shift $\tBAO$ at the
$\sim$\!0.5--1\% level per bin
\cite{Sanchez:2011,Nishimichi:2007,Crocce:2011}; photometric redshift
errors amplify these biases \cite{Padmanabhan:2007,Ross:2011}.  All 15
measurements share the SDSS footprint and its angular systematics
\cite{Ross:2011,Ho:2012,Leistedt:2013}: a coherent multiplicative
systematic would bias all bins in the same direction, producing the
observed one-signed offset.  If confirmed, the BAOtr-inclusive CPL
posteriors of Xu~et~al.~\cite{Xu:2026} --- showing strong Bayesian
evidence for dynamical dark energy --- would be unreliable, and the
DESI-only results (which are inconclusive) would be the more
trustworthy constraints. A third, independent probe points in the same direction.
Reconstructing $\DM/\rd$ from DES Type Ia supernovae at
$z = 0.510$ with a Planck $\rd$ yields
$\DM/\rd = 13.32 \pm 0.15$ (E.\ Ó Colgáin, private communication,
to appear in the revised version of Ref.~\cite{LopezHernandez:2025xxx}),
in agreement with DESI but discrepant with BAOtr at $\sim 3.2\sigma$;
a SH0ES-like $\rd$ would only strengthen this agreement. Further
support comes from an angular BAO analysis of the DESI DR1 BGS sample
by Ferreira~et~al.~\cite{Ferreira:2025}, which at $z \simeq 0.21$ and
$z \simeq 0.25$ finds mild preference for CPL over $\LCDM$
($1.5$--$2.6\sigma$ versus $2$--$3.3\sigma$ tension), with neither
model resolving the tension at these redshifts.

\subsubsection*{3D BAO systematics}

The DESI--SDSS consistency (Section~\ref{sec:robustness}) argues
strongly against this: both independent surveys give the same tension
pattern.  Only a shared theoretical assumption could affect both
identically.

\subsubsection*{New physics beyond CPL}

Models with non-smooth $w(z)$ --- $\Lambda_s$CDM
\cite{Akarsu:2020,Akarsu:2021,Akarsu:2023,Menci:2024rbq,
Akarsu:2025dmj,Akarsu:2024}, omnipotent dark energy \cite{Adil:2024},
composite dark energy with phantom matter
\cite{GomezValent:2025}, stochastic fluctuating $\Lambda$ from
quantum-gravity considerations \cite{Das:2023}, phantom-crossing scenarios
\cite{DiValentino:2021,Escamilla:2023,Alestas:2020zol,
Alestas:2020mvb,Marra:2021fvf}, or reconstructed scalar-tensor
gravity \cite{Efstratiou:2025iqi} (independently constrained by
NSBH gravitational-wave events \cite{Niu:2021})
--- could produce distance features
inaccessible to CPL.  However, the direct $z = 0.510$ clash
  ($3.7\sigma$ data-versus-data; $3.3\sigma$ against the best-fit
  $\LCDM$ prediction) cannot be resolved by modifying $\DM(z)$, since
  both datasets measure the same physical distance at the same
  redshift.\footnote{A caveat: the BAOtr shell has width
  $\Delta z = 0.02$ while the DESI LRG1 bin covers $0.4 < z < 0.6$.
  If $\DM(z)/\rd$ varies rapidly within this range, the effective
  redshifts could differ slightly.  However, the required $\sim$\!13\%
  change over $\Delta z \sim 0.05$ is implausible for any smooth
  expansion history.}  The model comparison alone contributes
  $\chi^2 \approx 11$ to the Method~A statistic, setting an
  irreducible floor.

Table~\ref{tab:explanations} summarises the three categories.

\begin{table}[ht]
\centering
\caption{Remaining explanations for the BAOtr--BAO\,3D tension.}
\label{tab:explanations}
\setlength{\tabcolsep}{4pt}
\footnotesize
\begin{tabular}{p{2.6cm}p{4.0cm}p{5.5cm}}
\toprule
Explanation & Key evidence & Discriminating test \\
\midrule
BAOtr systematics
  & One-signed offset; shared SDSS footprint;
    $z = 0.510$ direct clash
  & Angular BAO from DESI; DES/Euclid/LSST
    photometric BAO \\[6pt]
3D BAO systematics
  & DESI--SDSS consistency argues against
  & Alternative broadband parametrisations;
    blind mismatched-fiducial tests \\[6pt]
New physics beyond CPL
  & Cannot resolve $z = 0.510$ clash;
    could help at other~$z$
  & Fit $\Lambda_s$CDM and other models to
    combined data \\
\bottomrule
\end{tabular}
\end{table}

Of the three, BAOtr systematics are the most plausible given the
current evidence.

\subsection{Connection to the Hubble tension}
\label{sec:hubble}

The BAOtr--BAO\,3D tension is directly connected to the Hubble
tension.  BAO measurements constrain $\rd H_0$ through $\DM(z)/\rd$;
a dataset that prefers smaller $\DM/\rd$ at low redshift (BAOtr)
requires larger $H_0$ at fixed $\rd$.  Table~\ref{tab:H0} shows
this bifurcation.

\begin{table}[ht]
\centering
\caption{$H_0$ from different CMB+BAO combinations in CPL
\cite{Xu:2026}, with our $\chi^2_{\rm BAOtr}$
(Table~\ref{tab:chi2_summary}).  $T_{H_0}$ is relative to
$H_0^{\rm local} = 73.50 \pm 0.81$ \cite{Casertano:2025}.  The
$H_0$ values are from full MCMC \cite{Xu:2026}; $\chi^2_{\rm BAOtr}$
uses our DESI-optimized procedure.}
\label{tab:H0}
\setlength{\tabcolsep}{4.5pt}
\begin{tabular}{lccccr}
\toprule
Dataset & $w_0$ & $w_a$
  & $H_0$ [km\,s$^{-1}$\,Mpc$^{-1}$]
  & $T_{H_0}$ [$\sigma$]
  & $\chi^2_{\rm BAOtr}$ \\
\midrule
$\LCDM$
  & $-1.00$ & $0.00$
  & $67.31 \pm 0.49$ & $-6.6$
  & 41.6 \\[3pt]
CMB+SDSS
  & $-0.48$ & $-1.51$
  & $63.6^{+2.2}_{-2.5}$ & $-4.2$
  & 49.7 \\
CMB+DESI
  & $-0.42$ & $-1.75$
  & $63.9 \pm 2.0$ & $-4.4$
  & 49.8 \\[3pt]
CMB+PP\&SH0ES
  & $-0.69$ & $-1.70$
  & $70.87 \pm 0.68$ & $-2.5$
  & 37.4 \\
CMB+BAOtr
  & $-0.80$ & $-1.68$
  & $73.4^{+2.2}_{-3.8}$ & $-0.03$
  & ---$^a$ \\
\bottomrule
\multicolumn{6}{l}{\footnotesize $^a$Omitted because BAOtr data are
used in the fit itself, making $\chi^2_{\rm BAOtr}$ an in-sample} \\
\multicolumn{6}{l}{\footnotesize \phantom{$^a$}measure rather than an
independent tension statistic.} \\
\end{tabular}
\end{table}

The anti-correlation is clear: CMB+DESI yields $H_0 \approx 64$ with
$\chi^2_{\rm BAOtr} = 49.8$, while CMB+PP\&SH0ES yields
$H_0 \approx 71$ with $\chi^2_{\rm BAOtr} = 37.4$.  Within CPL,
reducing the BAOtr tension requires increasing $H_0$ (smaller
$\DM/\rd$), which pushes toward the local value but worsens the DESI
fit.  No single CPL model satisfies both.  This anti-correlation is a
structural limitation of the two-parameter CPL ansatz
\cite{Vagnozzi:2023nrq,Vagnozzi:2021tjv,Pantos:2026cxv,
Pedrotti:2024kpn}.

This is the CPL-specific expression of a more general, model-
independent result: the dataset-dependent late-time reconstructions
obtained by Gómez-Valent et al.~\cite{GomezValent:2024} show
that the bifurcation between low-$H_0$ (3D BAO) and high-$H_0$
(BAOtr) solutions persists even when no dark-energy parametrisation
is imposed.

\subsection{Limitations}
\label{sec:limitations}

Our DESI-optimization procedure determines $\Omega_m$ by minimising
$\chi^2_{\rm DESI}$ at each $(w_0, w_a)$ with $H_0$ set by
$\theta_*$.  A full MCMC with Planck likelihoods would also adjust
$\Omega_b h^2$ and $n_s$, potentially shifting $\Omega_m$ and $H_0$;
however, the close agreement between our parameters and those of
Xu~et~al.\ (Table~\ref{tab:Om_values}) confirms that the dominant
degeneracy is captured.

The BAOtr measurements are treated as statistically independent
despite sharing the SDSS angular systematics.  Correlated systematics
would reduce the effective degrees of freedom, weakening the quoted
$p$-values, though even halving the effective dof leaves the tension
significant.

Our analysis is restricted to CPL.  Models with more than two
dark-energy parameters or non-smooth $w(z)$ could improve the fit at
redshifts between the DESI anchors, but must confront the $z = 0.510$
direct clash ($\chi^2 \approx 11$), which sets a model-independent
floor.

We use only $\DM/\rd$ (and $\DV/\rd$ for BGS), neglecting $\DH/\rd$
and the intra-tracer correlations $\rho(\DM/\rd, \DH/\rd)$.
Including $\DH/\rd$ would tighten the DESI constraints but would not
affect the BAOtr comparison, which is sensitive only to $\DM/\rd$.

%% ═══════════════════════════════════════════════════════════════════
\section{Conclusions}
\label{sec:conclusions}
%% ═══════════════════════════════════════════════════════════════════

We have investigated whether the tension between the
fiducial-independent BAOtr dataset and the 3D BAO measurements from
DESI~DR2 and SDSS-IV can be attributed to the $\LCDM$ fiducial or
resolved within CPL.  Our main conclusions are:

\begin{enumerate}

\item \textbf{Published 3D BAO distances are fiducial-independent;
  rescaling overcorrects.}
  The fiducial cancels identically in $\alpha_\perp \times
  \DM^{\rm fid}/\rd^{\rm fid}$ (Eq.~\ref{eq:cancellation}), with
  residual effects at $\lesssim 0.3\%$.  Multiplying published
  distances by $\DM^{\rm CPL}/\DM^{\LCDM}$ overcorrects by a
  factor of 20--50 (Table~\ref{tab:overcorrection}).

\item \textbf{No CMB-consistent CPL model fits both DESI and BAOtr.}
  Scanning the $(w_0, w_a)$ plane with $\Omega_m$ and $H_0$
  determined by the $\theta_*$ constraint and DESI optimization, we
  find an inescapable trade-off: the best DESI-fitting models
  ($\chi^2_{\rm DESI} \lesssim 5$) give $\chi^2_{\rm BAOtr} \gtrsim
  42$, while reducing $\chi^2_{\rm BAOtr}$ to $\sim$\!37 requires
  $\chi^2_{\rm DESI} \gtrsim 8$ (Fig.~\ref{fig:tradeoff},
  Table~\ref{tab:chi2_summary}).

\item \textbf{The $z = 0.510$ bin sets an irreducible tension floor.}
  At this redshift, BAOtr gives $\DM/\rd = 11.91 \pm 0.42$ while
  DESI measures $13.59 \pm 0.17$ --- a $3.7\sigma$ data-versus-data
  clash.  Even against the best-fit $\LCDM$ prediction of $13.29$,
  the disagreement is $3.3\sigma$, contributing
  $\chi^2 \approx 11$ to the total.

\item \textbf{The results are robust.}
  The tension persists across two analysis methods (direct comparison
  and $\alpha$-interpolation), three extrapolation schemes, and
  substitution of SDSS for DESI ($\chi^2_B = 39.9$, $3.3\sigma$).

\item \textbf{The remaining explanations are BAOtr systematics (most
  likely) or new physics beyond CPL.}
  The one-signed offset, shared SDSS footprint, and direct
  $z = 0.510$ clash favour a coherent systematic in the BAOtr
  measurements.  New physics with non-smooth $w(z)$ could contribute
  at redshifts where the datasets do not overlap, but cannot resolve
  the direct clash.

\end{enumerate}

\medskip
\noindent\textbf{Outlook.}
The key next steps are:
\begin{itemize}
\item[(a)] Angular BAO analysis of DESI spectroscopic data,
  eliminating SDSS-specific systematics.
\item[(b)] Independent angular BAO from DES \cite{DES:2022},
  Euclid \cite{Euclid:2024}, and LSST \cite{LSST:2019}.
\item[(c)] Sub-percent reassessment of BAOtr projection effects,
  photometric-redshift biases, and angular systematics.
\item[(d)] Beyond-CPL models ($\Lambda_s$CDM, omnipotent dark energy,
  modified gravity) fitted to combined BAOtr + DESI + CMB data.
\item[(e)] Independent CPL constraints from third-generation
  gravitational-wave standard sirens \cite{Wang:2020} --- e.g.\ with
  Einstein Telescope and Cosmic Explorer --- offering sub-percent
  sensitivity to $(w_0, w_a)$ that bypasses BAO systematics entirely.
  
\end{itemize}

\noindent
The origin of the BAOtr--BAO\,3D tension remains open.  What our
analysis establishes is that the fiducial cosmology is not the answer,
and the CPL parametrisation cannot reconcile the two datasets.

\section*{Data availability}

The Python code to reproduce all figures and tables is publicly
available at:\\
\url{https://github.com/ipantos/BAOtr-BAO-3D-tension}.

\section*{Declaration of competing interest}

The authors declare no competing interests.

\section*{Acknowledgements}
The authors acknowledge networking support and participation in the CosmoVerse network.

\bibliographystyle{elsarticle-num}
\bibliography{references_new}

\end{document}